\documentclass[superscriptaddress,pre,aps,twocolumn,showpacs,floatfix]{revtex4-1}

\usepackage[dvips]{graphicx}
\usepackage{amsmath}
\usepackage{amssymb}
\usepackage[nice]{nicefrac}
\usepackage{color}
\usepackage{amsmath}
\usepackage{grffile}
\usepackage{times}

\newcommand{\sgn}{\mbox{sign}}

\newcommand{\Tau}{\mathcal{T}}

\begin{document}
\title{Escape from bounded domains driven by multi-variate $\alpha$-stable noises}

\author{Krzysztof Szczepaniec}
\email{kszczepaniec@th.if.uj.edu.pl}
\affiliation{Marian Smoluchowski Institute of Physics, and Mark Kac Center for Complex Systems Research, Jagiellonian University, ul. St. {\L}ojasiewicza 11, 30--348 Krak\'ow, Poland}

\author{Bart{\l}omiej Dybiec}
\email{bartek@th.if.uj.edu.pl}
\affiliation{Marian Smoluchowski Institute of Physics, and Mark Kac Center for Complex Systems Research, Jagiellonian University, ul. St. {\L}ojasiewicza 11, 30--348 Krak\'ow, Poland}

\date{\today}

\begin{abstract}
In this paper we provide an analysis of a mean first passage time problem of a random walker subject to a bi-variate $\alpha$-stable L\'evy type noise from a 2-dimensional disk. 
For an appropriate choice of parameters the mean first passage time reveals non-trivial, non-monotonous dependence on the stability index $\alpha$ describing jumps' length asymptotics both for spherical and Cartesian L\'evy flights.
Finally, we study escape from $d$-dimensional hyper-sphere showing that $d$-dimensional escape process can be used to discriminate between various types of multi-variate $\alpha$-stable noises, especially spherical and Cartesian L\'evy flights.
\end{abstract}

\pacs{
 05.40.Fb, 
 05.10.Gg, 
 02.50.-r, 
 02.50.Ey, 
 }
\maketitle

\section{Introduction\label{sec:introduction}}


Models of random walks \cite{montroll1965,montroll1984,metzler2000,shlesinger1983} hold a vital place in statistical physics as an universal tool for
large amount of physical systems \cite{metzler2004,Bartumeus2005,esposito2008,scalas2006,abad2010,barhaim1998}.
The canonical, well understood, Brownian motion still plays
the major role, however various non-Gaussian and non-Markovian generalizations have been introduced.
Heavy tailed fluctuation have been observed in versatilities of models \cite{solomon1993,solomon1994,chechkin2002b,boldyrev2003}
including physics, chemistry or biology \cite{shlesinger1995,nielsen2001},
paleoclimatology \cite{ditlevsen1999b} or economics
\cite{mantegna2000} and epidemiology \cite{brockmann2006,dybiec2009c} to name a few.
Observations of the so-called L\'evy flights boosted the theory of random walks and noise induced phenomena into new directions \cite{chechkin2004,sokolov2003,dubkov2008,rypdal2010,barthelemy2008,pasternak2009,lomholt2005,klages2008,srokowski2009b,srokowski2009b,dubkov2013} which involve examination of space fractional diffusion equation (Smoluchowski-Fokker-Planck equation) and stimulated development of more general theory \cite{janicki1994,samorodnitsky1994}. Theory of $\alpha$-stable processes allows for examination of more general fluctuations than Gaussian including them as a limiting case. The high efficiency and generality of $\alpha$-stable processes is based on the generalized central limit theorem \cite{gnedenko1968,meerschaert2001}, which provides extension of the standard central limit theorem to the situation when assumption of finite variance of elements is relaxed.
Consequently, $\alpha$-stable processes provide natural tool for description of systems revealing power-law fluctuations.

The problem of noise-induced escape from finite intervals or semi-infinite domains, with the canonical example of one-dimensional diffusion, has been studied in great details in various non-Gaussian and non-Markovian \cite{Benichou2005a,zoia2007,dybiec2010,majumdar2010,dybiec2010c,bertoin1996first,garcia2012,demulatier2013} regimes including symmetric and asymmetric $\alpha$-stable L\'evy type noise as an especially important extension.
Analysis of the mean first passage time (escape time) from a finite interval provides an insight into the complexity of stable noise.
The interplay of noise parameters and non-locality of boundary conditions \cite{dybiec2006,zoia2007} result in reach behavior of $\alpha$-stable noise driven systems.
In particular, their non-triviality is manifested by the failure of the method of images \cite{chechkin2003b}, leapovers \cite{koren2007,koren2007b}, non-trivial properties of stationary states \cite{chechkin2002,chechkin2003,chechkin2004,dybiec2007d,srokowski2010,dubkov2007,sliusarenko2012}  and non-linear, non-monotonous behavior of the mean first passage \cite{dybiec2006,zoia2007} time which measures efficiency of the noise facilitated escape.

So far majority of research focuses mainly on uni-variate processes. Extensions into multi-variate domains seem natural and well defined,
yet still challenging due to their non-triviality. General approach to 2-dimensional L\'evy flights assume that a step direction is chosen from uniform distribution on a circle and jumps' lengths are distributed according to heavy-tailed densities \cite{teuerle2009,chechkin2002b} what assures isotropic probability of finding a random walker at a given distance from the starting point. Due to the generalized central limit theorem, such approach leads to desired spherical L\'evy flights, providing effective model for various processes \cite{Edwards2007}. Nevertheless, an alternative and natural approach based on bi-variate $\alpha$-stable distributions has been suggested \cite{teuerle2009}.
In such a case, on the one hand, the whole process is determined by a multi-variate $\alpha$-stable density which contains all the information about increments of the process. On the other hand plenitude of possible spectral measures, leading to various (fractional) diffusion equations, shifts the main difficulties into a different place than approach formerly applied \cite{blumenthal1961, *getoor1961, *kac1950distribution, *widom1961stable, *kesten1961random}.

In this paper we explore a 2-dimensional escape problem of a random walker driven by a bi-variate L\'evy stable noise
which provides a natural, yet non-trivial, extension of 1D $\alpha$-stable noises to higher dimensions.
Starting from known results for 1D escape problem, with the non-monotonous dependence of the mean first passage time as a function of the stability index $\alpha$, see \cite{dybiec2006,zoia2007}, 
we search for analogous behavior in the noise driven escape process from a disk.
In the very limited number of cases we compare results of numerical simulations with exact results 
\cite{blumenthal1961, *getoor1961, *kac1950distribution, *widom1961stable, *kesten1961random,redner2001,borodin2002}.
Finally, we compare spherical $\alpha$-stable motions (spherical L\'evy flights) and so-called Cartesian L\'evy flights.

\section{Model and results\label{sec:model}}


\subsection{1D motivation\label{sec:1d}}

A motion of a free particle subject to the symmetric $\alpha$-stable L\'evy type noise is described by the Langevin equation
\begin{equation}
 \frac{dx}{dt} = \sigma \zeta_{\alpha}(t),
 \label{eq:langevin}
\end{equation}
which can be rewritten as
$
 dx = \sigma dL_{\alpha}(t),
 \label{eq:langevin2}
$
where $L_{\alpha}(t)$ is a symmetric $\alpha$-stable motion \cite{janicki1996} i.e. stochastic process with independent increments distributed according to an $\alpha$-stable density \cite{samorodnitsky1994,janicki1996}.
$\zeta_{\alpha}(t)$ represents a white $\alpha$-stable noise which is a formal time derivative of a symmetric $\alpha$-stable motion.
The characteristic function of symmetric $\alpha$-stable densities is  \cite{samorodnitsky1994,janicki1994}
\begin{equation}
 \phi(k)= \mathbb{E} \left[ e^{ikX} \right] = 
 \exp\left[ -\sigma^\alpha |k|^\alpha   \right] 
 \label{eq:characteristic1d}
\end{equation}
where $\alpha \in (0,2]$ is the stability index, $\sigma > 0$ is the scale parameter. 
For $\alpha<2$, symmetric $\alpha$-stable densities have the power-law asymptotics of $|x|^{-(\alpha+1)}$ type 
resulting in divergence of fractional moments $\langle |x|^\nu \rangle$ of order $\nu>\alpha$.

Closed formulas for probability densities corresponding to the characteristic function~(\ref{eq:characteristic1d}) are known only in a limited number of cases. For $\alpha=2$, the Gaussian distribution
$
 f(x)= \exp\left[ -{x^2}/{(4\sigma^2)} \right]/\sqrt{4\pi\sigma^2}
$
is recovered, which is the only one $\alpha$-stable density possessing all moments.
For $\alpha=1$, the Cauchy density
$
f(x)=\sigma /\left[ \pi ( {x^2+\sigma^2}) \right]
$
is recovered with the mean value defined as the principal value of the appropriate integral.
In the most general realms, $\alpha$-stable densities can be asymmetric and shifted. In such a case the characteristic function~(\ref{eq:characteristic1d}) depends on an additional asymmetry parameter $\beta$ and the location parameter $\mu$, see \cite{samorodnitsky1994,janicki1994}.

In 1D, Eq.~(\ref{eq:langevin}) can be associated with the following (space-fractional) Smoluchowski-Fokker-Planck equation \cite{fogedby1994,metzler1999,yanovsky2000,schertzer2001,dubkov2008}
\begin{equation}
\frac{\partial p(x,t|x_0,0)}{\partial t}=\sigma^\alpha\frac{\partial^\alpha p(x,t|x_0,0)}{\partial |x|^\alpha}.
\label{eq:ffpe}
\end{equation}
In the above equation the Riesz-Weyl fractional derivative is defined by the Fourier transform, i.e. $\mathcal{F}\left[ \frac{\partial^\alpha f(x)}{\partial |x|^\alpha} \right]=-|k|^\alpha\mathcal{F}\left[f(x)\right]$.
For $\alpha=2$, any L\'evy stable noise is equivalent to the Gaussian white noise (with the standard deviation $\sqrt{2}\sigma$, see \cite{janicki1994}) and the fractional Smoluchowski-Fokker-Planck equation~(\ref{eq:ffpe}) takes its standard form, i.e. $\partial^\alpha/\partial |x|^\alpha \to \partial^2/\partial x^2$.
Solutions of the fractional diffusion equation~(\ref{eq:ffpe}) for $p(x,0|0,0)=\delta(x)$ are given by symmetric $\alpha$-stable densities with time dependent scale parameter $\sigma(t) \propto t^{1/\alpha}$, see Eq.~(\ref{eq:characteristic1d}).
In the restricted space, due to presence of boundaries, except $\alpha=2$, usually it is not possible to find formulas for the density $p(x,t|0,0)$, see below.

Motion of the particle can be restricted by geometric constraints which introduce boundary conditions.
For example, a domain of motion can be a finite $[-L,L]$ interval restricted by two absorbing boundaries.
Presence of absorbing boundaries require special care. In particular, 
for $\alpha<2$, trajectories of $\alpha$-stable motions are discontinuous.
Consequently, for $\alpha<2$, fractional Smoluchowski-Fokker-Planck equation is associated with the non-local boundary conditions, i.e.
$p(x,t|x_0,0)=0$ for $|x|\geqslant L$, see \cite{dybiec2006,zoia2007} due to leapovers \cite{koren2007,koren2007b}.
In the Gaussian case ($\alpha=2$) boundary conditions are local (defined at $|x|=L$ only) and the solution of Eq.~(\ref{eq:ffpe})
can be constructed using method of images \cite{cox1965,borodin2002,redner2001} or Fourier series \cite[Eq.~(81)]{cox1965}.
For an extended discussion see \cite{dybiec2012fractional}.
From $p(x,t|x_0,0)$ one can calculate the mean first passage time $\Tau$, i.e. the average time when the particle leaves the domain of motion for the first time $\tau$. In particular, for $\alpha=2$
\begin{equation}
\Tau = \langle \tau \rangle = \int_0^\infty dt  \int_{|x|\leqslant L} p(x,t|x_0,0)dx  = \frac{L^2}{2{\sigma}^2}.
\label{eq:mfpt-sum-1d}
\end{equation}
The mean first passage time $\Tau = \langle \tau \rangle$ from a bounded interval is finite, regardless of $\alpha$, see below.

Alternatively, the mean first passage time $\Tau=\langle \tau \rangle$, see Eq.~(\ref{eq:mfpt-sum-1d}), at which a particle leaves the region $\Omega=[-L,L]$ for the first time can be directly calculated from the backward Smoluchowski-Fokker-Planck equation \cite{redner2001,gardiner2009}, which for $\alpha=2$ has the form
\begin{equation}
 \sigma^2 \frac{\partial^2 \Tau(x)}{\partial x^2}   =  -1,
 \label{eq:backward}
\end{equation}
with the boundary condition $\Tau|_{\partial \Omega}=0$ and the initial condition $x(0) \in \Omega$. The MFPT calculated from Eq.~(\ref{eq:backward}) is given by Eq.~(\ref{eq:mfpt-sum-1d}), i.e. $\Tau(0)=\frac{L^2}{2\sigma^2}$.
Eq.~(\ref{eq:backward}) can be easily generalized to the fractional case \cite{zoia2007}
\begin{equation}
\sigma^\alpha \frac{\partial^\alpha \Tau(x)}{\partial |x|^\alpha}=-1,
 \label{eq:backward-fractional}
\end{equation}
or into higher dimensions $d$. 
The generalization of Eq.~(\ref{eq:backward}) into Eq.~(\ref{eq:backward-fractional}) affects  boundary conditions, which become non-local. Consequently, the whole exterior of $\Omega$ is absorbing because a particle can escape from the domain of motion $\Omega$ without hitting the boundary.

For $\alpha<2$,
the mean first passage time for the escape from a finite interval $[-L,L]$ restricted by two absorbing boundaries can be calculated for any value of the stability index $\alpha$ \cite{zoia2007}.
For $x(0)=0$ the formula for the mean first passage time reads
\begin{equation}
 \Tau = \langle \tau \rangle = \frac{L^\alpha}{\Gamma(1+\alpha)\sigma^\alpha}.
 \label{eq:mfpt-1d}
\end{equation}
For $\alpha=2$, Eq.~(\ref{eq:mfpt-1d}) reduces to Eq.~(\ref{eq:mfpt-sum-1d}).
Additionally, as for $\alpha=2$, the first passage time density has exponential asymptotics \cite{dybiec2009g,dybiec2010c,dybiec2014estimation} what is a typical property of Markovian escape processes.

\begin{figure}[!ht]
 \includegraphics[angle=0,width=0.85\columnwidth]{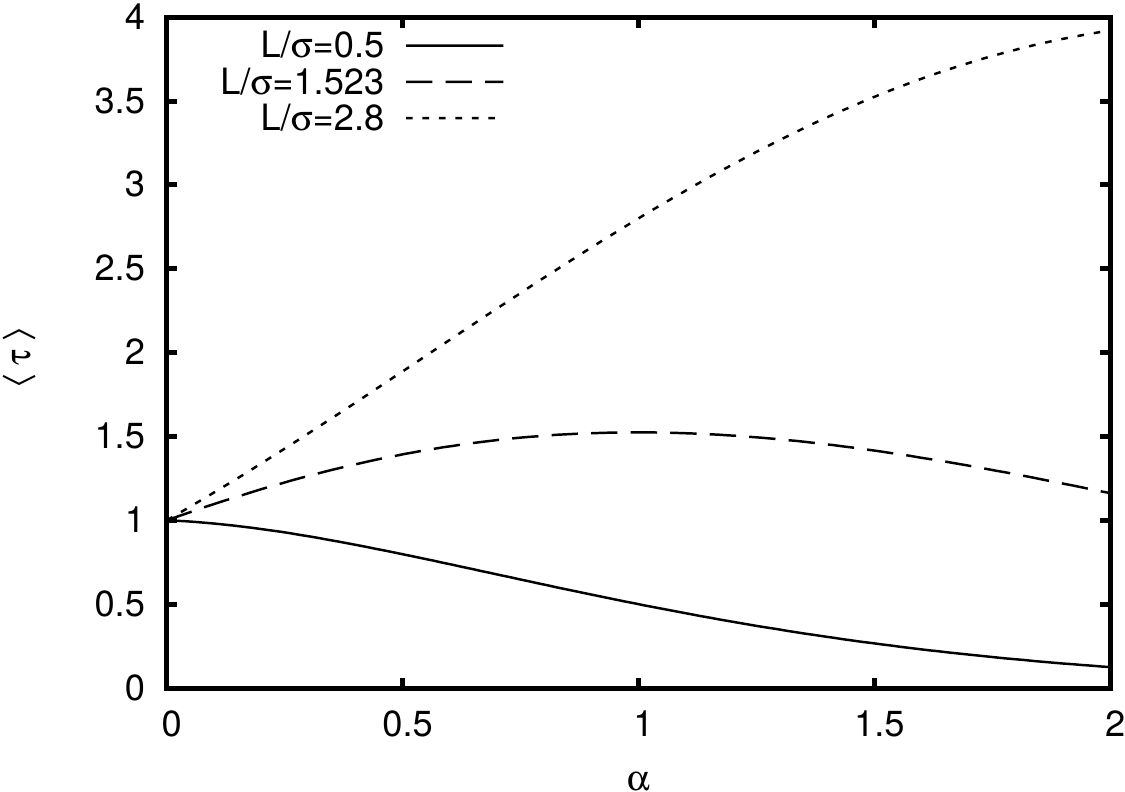}
 \caption{The mean first passage time $\Tau = \langle  \tau \rangle$, i.e. Eq.~(\ref{eq:mfpt-1d}), for various values of $L/\sigma$.
 $L/\sigma=1.523$  assures that the MFPT is maximal for $\alpha \approx 1$.
 For $L/\sigma$ large enough the maximal mean first passage time is located at $\alpha=2$ while for low values of $L/\sigma$   is reached for $\alpha \approx 0$.
 }
 \label{fig:approximation}
\end{figure}

The extrema of $\Tau = \langle \tau \rangle$ as a function of the stability index $\alpha$ can be at boundaries of the stability index range ($0<\alpha \leqslant 2$).
Nevertheless, the most interesting is the possibility of observing maximal value of the MFPT for an intermediate $\alpha$.
Differentiating Eq.~(\ref{eq:mfpt-1d}) with respect to $\alpha$ one can find an approximate relation between the interval half-width $L$ and the scale parameter $\sigma$ for which the maximal MFPT is recorded at given (fixed) $\alpha$. For example, assuming that the MFPT is maximal for $\alpha=1$ one approximately gets
$
 L/\sigma \approx 1.523.
$
This demonstrates that for a given interval half-width $L$  it is possible to find such a scale parameter $\sigma$ that the MFPT depends in a non-monotonous way on the stability index $\alpha$. Fig.~\ref{fig:approximation} demonstrates the dependence of the mean first passage time~(\ref{eq:mfpt-1d}) on the stability index $\alpha$ for various values of $L/\sigma$. Changes in $L/\sigma$ shift location of maximum of the MFPT from $\alpha\approx 0$ (small $L/\sigma$) to $\alpha=2$ (large $L/\sigma$).
In particular, for $L/\sigma\approx 1.523$ maximal value of the mean first passage time is recorded for $\alpha \approx 1$.

\begin{figure}[!ht]
 \includegraphics[angle=0,width=0.85\columnwidth]{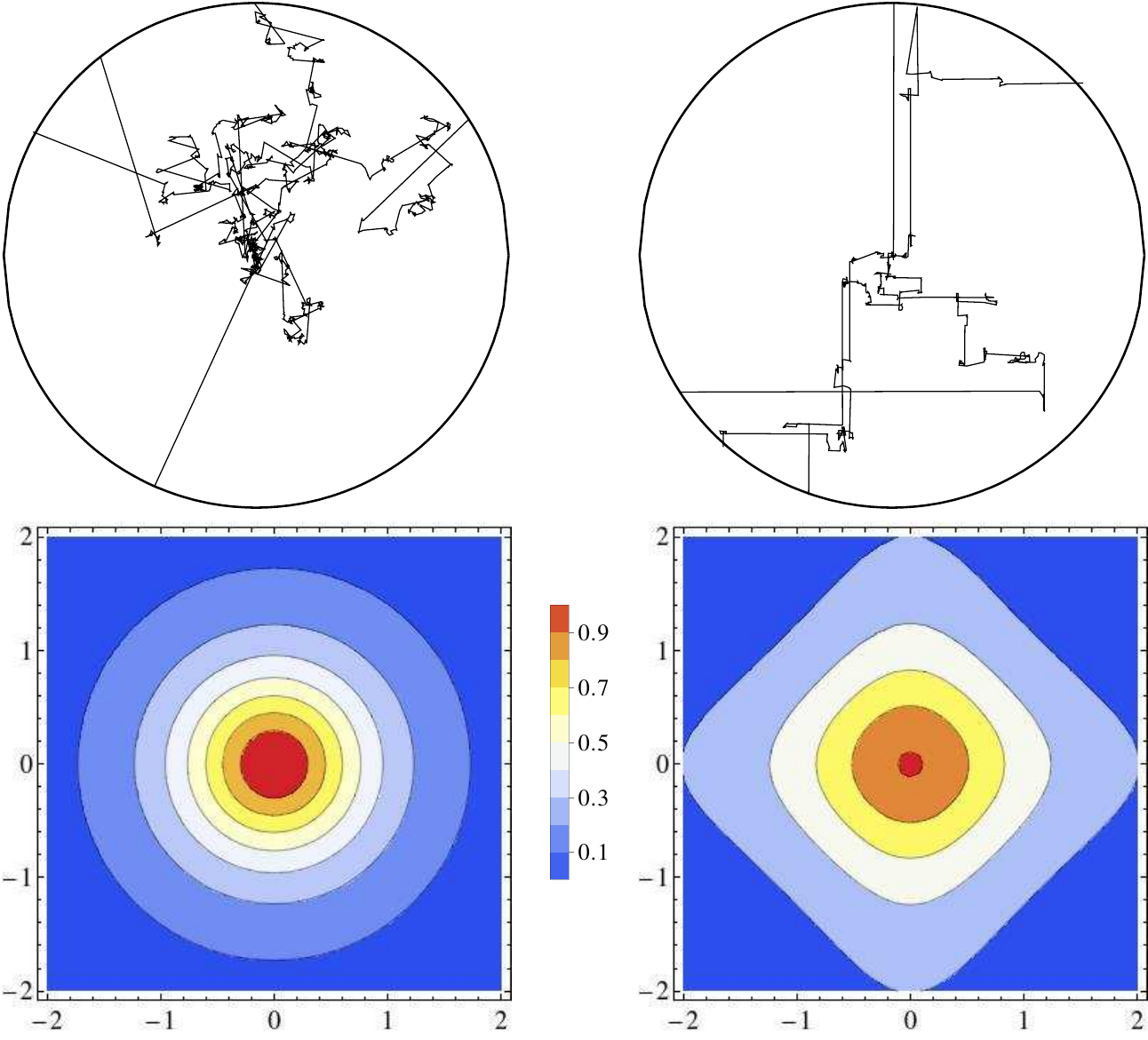}
 \caption{(Color online) Five trajectories of spherically symmetric 2D $\alpha$-stable motions (top left panel) and Cartesian (Cauchy) L\'evy flights (top right panel). 
  Spherical (bottom left panel) and Cartesian (bottom right panel) Cauchy distributions, see Eqs.~(\ref{eq:scauchy}) and (\ref{eq:kcauchy}).
 }
 \label{fig:trajectories}
\end{figure}

\subsection{Escape in 2D\label{sec:2d}}

Let us consider a 2D motion of a free overdamped particle subject to the bi-variate $\alpha$-stable L\'evy type noise
\begin{equation}
 \frac{d\boldsymbol{x}}{dt} = \sigma \boldsymbol{\zeta}_{\alpha}(t).
 \label{eq:langevin2d}
\end{equation}
Analogously like in 1D, Eq.~(\ref{eq:langevin2d}) can be rewritten in the incremental form
$
 d\boldsymbol{x} = \sigma d\boldsymbol{L}_{\alpha}(t),
 \label{eq:langevin2d2}
$
where $\boldsymbol{L}_{\alpha}(t)$ is a bi-variate $\alpha$-stable motion \cite{samorodnitsky1994}.
As in 1D, the bi-variate $\alpha$-stable noise is a formal time derivative of the bi-variate $\alpha$-stable motion.
Therefore, increments of $\boldsymbol{L}_{\alpha}(t)$ are independent and distributed according to the bi-variate $\alpha$-stable density. General $d$-variate $\alpha$-stable densities have the characteristic function
\begin{widetext}
\begin{equation}
 \phi(\boldsymbol{k}) =
 \left\{
 \begin{array}{lcl}
 \exp\left\{ -\int_{S_d} |\langle \boldsymbol{k},\boldsymbol{s} \rangle|^\alpha \left[ 1 -i\sgn(\langle \boldsymbol{k},\boldsymbol{s} \rangle)\tan\frac{\pi\alpha}{2} \right]\Lambda(d\boldsymbol{s}) +i \langle \boldsymbol{k},\boldsymbol{\mu}^0 \rangle \right\} & \mbox{for} & \alpha\neq 1,\\
 \exp\left\{ -\int_{S_d} |\langle \boldsymbol{k},\boldsymbol{s} \rangle| \left[ 1 +i\frac{2}{\pi}\sgn(\langle \boldsymbol{k},\boldsymbol{s} \rangle)\ln(\langle \boldsymbol{k},\boldsymbol{s} \rangle) \right]\Lambda(d\boldsymbol{s}) +i \langle \boldsymbol{k},\boldsymbol{\mu}^0 \rangle \right\} & \mbox{for} & \alpha = 1,
 \end{array}
 \right.
 \label{eq:characteristicdd}
\end{equation}
\end{widetext}
where $\langle \boldsymbol{k} , \boldsymbol{s} \rangle$ represents the scalar product, $\Lambda(\cdot)$ stands for the spectral measure on the unit sphere $S_d$ of $\mathbb{R}^d$ and $\boldsymbol{\mu}^0$ is a vector in $\mathbb{R}^d$, see \cite{samorodnitsky1994}.
Bi-variate case corresponds to $d=2$.
The spectral measure $\Lambda(\cdot)$ replaces skewness and scale parameters ($\beta$ and $\sigma$) which characterize 1D $\alpha$-stable densities. Multi-variate $\alpha$-stable density is said to be symmetric if the spectral measure is symmetric, see \cite{samorodnitsky1994,teuerle2009,teuerle2012}.
The multi-variate $\alpha$-stable motion $\boldsymbol{L}_{\alpha}(t)$ can be generated in analogous way like $L_{\alpha}(t)$, see \cite{chambers1976,weron1996,janicki1994}. The only difference is in the approximation scheme which relies on generation of multi-variate $\alpha$-stable random variables that can be generated by general methods described in \cite{modarres1994method,nolan1998b,samorodnitsky1994}.

The 2D case significantly differs from 1D, because bi-variate $\alpha$ stable noises are determined by the spectral measure $\Lambda(\cdot)$. Various choices of spectral measures result in different escape scenarios and different (fractional) diffusion equations \cite{samko1993,chechkin2002b,szczepaniec2014stationary}. Here, we focus on the uniform continuous spectral measures resulting in spherical $\alpha$-stable motions (spherical L\'evy flights) and uniform discrete spectral measures concentrated on intersections of the unit circle (or hyper-sphere) with axes leading to the so called Cartesian L\'evy flights \cite{vahabi2013,samorodnitsky1994,chechkin2000b,chechkin2002b}. 
The uniform continuous spectral measure $\Lambda(\cdot)$
corresponds to the situation when $\alpha$-stable densities are spherically symmetric, i.e. their isolines are circles ($d=2$), spheres ($d=3$) or hyper-spheres ($d>3$).
In such a case the Langevin equation~(\ref{eq:langevin2d}) is associated with the following fractional diffusion equation \cite{yanovsky2000,schertzer2001,samko1993}
\begin{eqnarray}
 \frac{\partial p(\boldsymbol{x},t|\boldsymbol{x}_0,t)}{\partial t} & = - \sigma^\alpha (-\Delta)^{\alpha/2} p(\boldsymbol{x},t|\boldsymbol{x}_0,t),
\label{eq:ffpemd}
\end{eqnarray}
where $-(-\Delta)^{\alpha/2}$ is the fractional Riesz-Weil derivative (laplacian) defined via its Fourier transform \cite{samko1993}
 \begin{equation}
 \mathcal{F}\left[ -(-\Delta)^{\alpha/2} p(\boldsymbol{x},t|\boldsymbol{x}_0,t) \right] = -|\boldsymbol{k}|^\alpha \mathcal{F}\left[ p(\boldsymbol{x},t|\boldsymbol{x}_0,t) \right].
 \label{eq:weil}
 \end{equation}
For the discrete uniform spectral measure located on intersections of the unit sphere with the axes the fractional Smoluchowski-Fokker-Planck equation takes the form
\begin{equation}
 \frac{\partial p(\boldsymbol{x},t|\boldsymbol{x}_0,t)}{\partial t}  = \sigma^\alpha \left[ \frac{\partial^\alpha}{\partial |x|^\alpha} + \frac{\partial^\alpha}{\partial |y|^\alpha} \right] p(\boldsymbol{x},t|\boldsymbol{x}_0,t),
\label{eq:ffpekart}
 \end{equation}
 see Eq.~(\ref{eq:ffpe}).
 The associated backward equation~(\ref{eq:backward-fractional}) transforms into multi-dimensional domain in the same manner like the forward equations~(\ref{eq:ffpemd}) and~(\ref{eq:ffpekart}).
 Analogously like in 1D, Eqs.~(\ref{eq:ffpemd}) and~(\ref{eq:ffpekart}) are associated with the non-local boundary conditions, i.e. whole exterior of the domain of motion is absorbing.
 The MFPT can be calculated either from backward diffusion equation, diffusion equation (see Eqs.~(\ref{eq:ffpemd}),~(\ref{eq:ffpekart}) and~(\ref{eq:mfpt-sum-1d})) or Monte Carlo simulations (Langevin dynamics), which is the main methodology applied within the current presentation.

 The difference between 2D $\alpha$-stable motions, e.g. spherical and Cartesian L\'evy flights, is especially well visible for $\alpha=1$. In 2D, $\alpha$-stable processes lead to Cauchy distributions. For the uniform continuous spectral measure the (isotropic) radial Cauchy distribution is recovered
 \begin{equation}
  p(x,y)=\frac{1}{2\pi}\frac{\sigma}{(x^2+y^2+\sigma^2)^{3/2}}.
  \label{eq:scauchy}
 \end{equation}
 For the discrete uniform spectral measure located on intersections of the unit sphere with axes the probability is a product of two 1D Cauchy densities
 \begin{equation}
  p(x,y)=\frac{1}{\pi}\frac{\sigma}{x^2+\sigma^2} \times \frac{1}{\pi}\frac{\sigma}{y^2+\sigma^2}.
  \label{eq:kcauchy}
 \end{equation}
 In both cases the scale parameter grows like $\sigma \propto t$.
 In the limit of $\alpha=2$, spherical and Cartesian L\'evy flights tend to bi-variate Brownian motion demonstrating that both scenarios are equivalent for $\alpha=2$.
 Trajectories of spherical and Cartesian (Cauchy) L\'evy flights with $\alpha=1$ are presented in top left and top right panels of Fig.~\ref{fig:trajectories}. Spherical L\'evy flights are isotropic, while Cartesian L\'evy flights display preference to horizontal and vertical jumps.
 Both types of Cauchy densities are presented in bottom panel of Fig.~\ref{fig:trajectories}.

This time as a finite domain of motion a disk of radius $R$ is considered. The absorbing boundary is defined by the disk edge.
If the trajectory crosses the disk edge the particle is removed from the domain of motion and the first passage time $\tau=t$ is recorded. 
Such an approach guarantees proper implementation of boundary conditions because the whole exterior of $\Omega$ is absorbing.

The main aim is to check if non-monotonous dependence of the MFPT on the stability index $\alpha$ can be observed in higher dimensional systems in analogous way like it was observed in 1D, see Eq.~(\ref{eq:mfpt-1d}) and~Fig.~\ref{fig:approximation}.
In order to verify this hypothesis we use extensive numerical simulations.
Main simulations were performed on a circle with radius $R=5$, number of repetitions $N=10^5$ and the integration time step $\Delta t=10^{-4}$.
Initially a random walker was located in the center of the disk, i.e. $\boldsymbol{x}(0)=\boldsymbol{0}$.
The first passage time $\tau$ is estimated as
\begin{equation}
 \tau = \min\{t>0 \;\;:\;\; \boldsymbol{x}(0)=\boldsymbol{0} \mbox{ and }  \boldsymbol{x}(t) \notin \Omega \},
 \label{eq:mfpt-trajectory}
\end{equation}
where $\Omega$ is the domain of motion, e.g. interval, disk, etc.
The mean first passage time $\Tau=\langle \tau \rangle$ is the average first passage time $ \tau$.

For $\alpha=2$, i.e. for the bi-variate Gaussian white noise, using Eq.~(\ref{eq:backward}) it is possible to calculate the mean first passage time $\Tau$ exactly.
Due to the system symmetry, the only relevant variable is the initial distance from the disk center $r$.
Rewriting Eq.~(\ref{eq:backward}) in polar coordinates one gets
\begin{equation}
 \frac{d^2 \Tau(r)}{d r^2} + \frac{1}{r} \frac{d \Tau(r)}{d r}=-\frac{1}{\sigma^2},
 \label{eq:backward-polar}
\end{equation}
with the boundary condition $\Tau(R)=0$.
The solution of Eq.~(\ref{eq:backward-polar}) is
\begin{equation}
 \Tau(r)=\frac{1}{4\sigma^2}\left[ R^2- r^2 \right],
 \label{eq:2dmfpt}
\end{equation}
which for $r=0$ reduces to $R^2/4\sigma^2$, i.e. the MFPT from the disk of radius $R$ is two times smaller than the MFPT from the interval of the half-width $R$.

\begin{figure}[!ht]
 \includegraphics[angle=0,width=0.95\columnwidth]{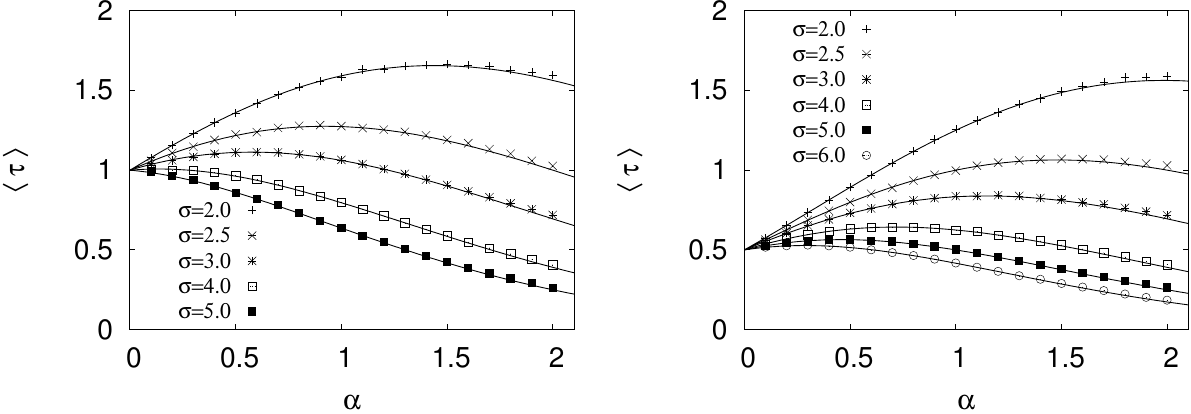}
 \caption{The mean first passage time for the escape from the (2D) disk as a function of the stability index $\alpha$ for spherical (left panel) and Cartesian (right panel) L\'evy flights. The whole disk edge is absorbing. A particle starts its diffusive motion in the disk center. The disk radius is $R=5$. Various curves correspond to various values of 
 the scale parameter $\sigma$. Solid lines in the left panel present exact values of the MFPT given by Eq.~(\ref{eq:general-mfpt}).
 Solid lines in the right panel present the MFPT given by Eq.~(\ref{eq:mfptclf}).
 Error bars are smaller than the symbol size.
 }
 \label{fig:mfpt-full-circle}
\end{figure}

For the disk with entire absorbing edge, the mean first passage time is the time in which random walker, starting at the center of the disk
reaches or passes over the edge of the disk for the first time. While, in general the behavior of the MFPT depends on the noise parameters, it is possible
to find such a range of scale parameter $\sigma$ for which the MFPT becomes a non-monotonous function of the stability index $\alpha$. 
Also, the position of MFPT maxima shifts with the change in the scale parameter $\sigma$. Fig.~\ref{fig:mfpt-full-circle} demonstrates dependence of the mean first passage time from the disk on the stability index $\alpha$
for spherical (left panel) and Cartesian (right panel) L\'evy flights.
Various curves correspond to various values of the scale parameter $\sigma$.

Non-monotonous dependence of the mean first passage time, as a function of the stability index $\alpha$, is observed for quite narrow range of the scale parameter $\sigma$, see Fig.~\ref{fig:approximation}.
Nevertheless, this special range increases with the increase of the system size.
For low $\sigma$, the MFPT monotonically increases with the stability index $\alpha$.
For intermediate $\sigma$, non-monotonous dependence of the MFPT is observed.
With the further increase of the scale parameter $\sigma$ the MFPT monotonically decreases as a function of the stability index $\alpha$.
The behavior of the MFPT is a consequence of the interplay between the stability index $\alpha$ and the scale parameter $\sigma$.
The stability index $\alpha$ not only controls the tail asymptotics of $\alpha$-stable densities but also affects its width, as measured by the inter quantile distance.
For Cartesian L\'evy flights the non-monotonous dependence of the mean first passage time is observed for slightly different values of scale parameters than for spherical L\'evy flights, compare left and right panels of Fig.~\ref{fig:mfpt-full-circle}.

From formulas~(\ref{eq:mfpt-1d}), (\ref{eq:2dmfpt}) and dimensional analysis one can predict that the mean first passage time scales with the scale parameter as
\[
 \Tau \propto \frac{1}{\sigma^\alpha}.
\]
Such a scaling is very well confirmed by computer simulations (results not shown) and 
the general formulas (\ref{eq:general-mfpt}) and~(\ref{eq:mfptclf}), see below.
Finally, in the all cases the first passage time density has exponential tails, what is the general property of continuous time and space Markov escape process from finite domains even for processes with discontinuous trajectories, i.e. $\boldsymbol{x}(t)$ with $\alpha<2$.

The non-monotonous dependence of the mean first passage time, can be also observed for spherical L\'evy flights scheme in which step lengths are drawn from symmetric $\alpha$-stable density and jump directions are uniformly distributed \cite{teuerle2009,chechkin2002b,vahabi2013} (results not shown). Due to the generalized central limit theorem such spherical L\'evy flights converge to the isotropic bi-variate $\alpha$-stable motions.

\subsection{Escape in $d$-dimensions\label{sec:dd}}

\begin{figure}[!ht]
\includegraphics[angle=0,width=0.95\columnwidth]{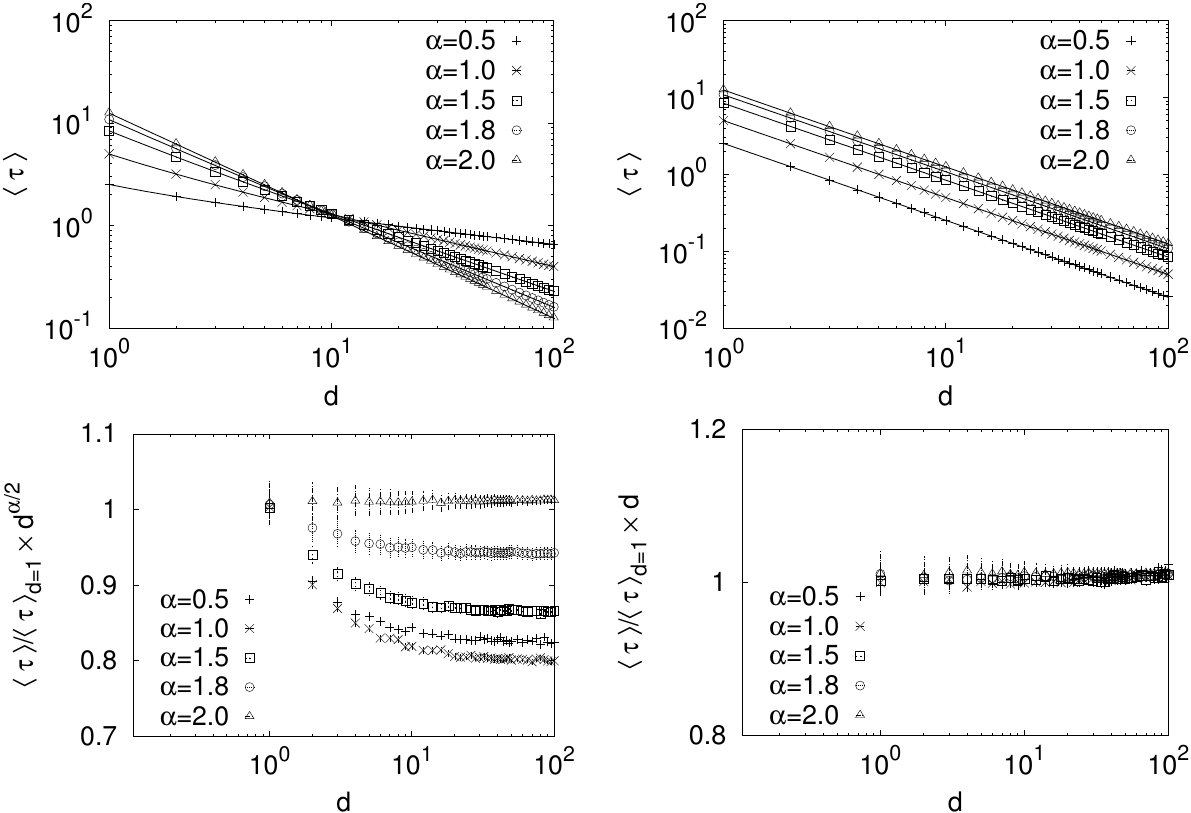}
 \caption{The mean first passage times $\langle \tau \rangle$ for the escape from the $d$-dimensional hyper-sphere (top row) and
 ratio of MFPTs for $d$-dimensional hyper-sphere and 1D sphere (interval), i.e. $\langle \tau \rangle / \langle \tau \rangle_{d=1} \times d^{\alpha/2}$ (left bottom panel) for spherical L\'evy flights (left column) and $\langle \tau \rangle / \langle \tau \rangle_{d=1} \times d$ (right bottom panel) for Cartesian L\'evy flights (right column). Hyper-sphere radius $R=5$.
 Various curves correspond to various values of the stability index $\alpha$.
 Lines in the left top panel show theoretical values of the MFPT given by Eqs.~(\ref{eq:mfpt-d-spere}),~(\ref{eq:general-mfpt}) and~(\ref{eq:mfptclf}).
 Error bars in the top panel are smaller than the symbol size.
}
 \label{fig:hypersphere}
\end{figure}

For $\alpha=2$, in $d$-dimensions, the MFPT can be calculated by use of Eq.~(\ref{eq:backward}), which in the polar coordinates takes the form
\begin{equation}
 \Tau''(r)+\frac{d-1}{r}\Tau'(r)=-\frac{1}{\sigma^2}
 \label{eq:backward-dspherical}
\end{equation}
and has the solution
\begin{equation}
 \Tau(r) =\frac{1}{2\sigma^2d}\left[ R^2- r^2 \right],
 \label{eq:mfpt-d-spere}
\end{equation}
which for $r=0$ reduces to $R^2/2d \sigma^2 $, i.e. the MFPT is equal to $\langle \tau \rangle_{d=1}/d$.
The solution~(\ref{eq:mfpt-d-spere}) perfectly agrees with results of numerical simulations, see Fig.~\ref{fig:hypersphere}, where not only results of numerical simulations for $\alpha=2$ are presented but also theoretical values given by Eqs.~(\ref{eq:mfpt-d-spere}), (\ref{eq:general-mfpt}) and (\ref{eq:mfptclf}), see below.

\begin{figure}[!ht]
\includegraphics[angle=0,width=0.95\columnwidth]{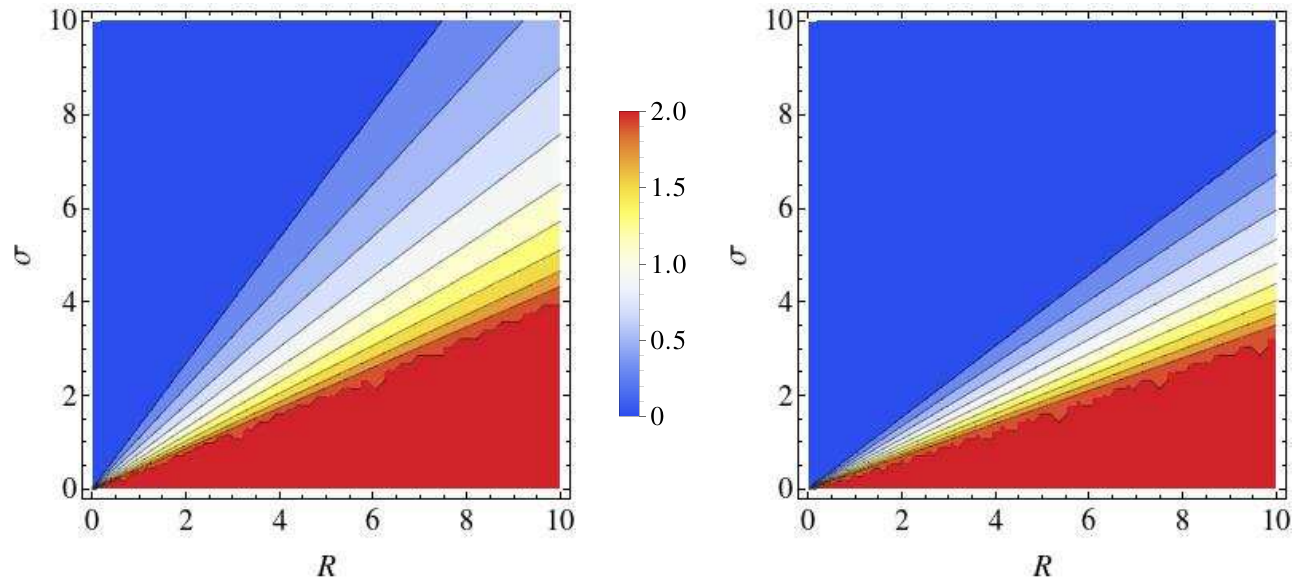}
 \caption{
 (Color online) Values of the stability index $\alpha$ leading to the maximal mean first passage time for spherical L\'evy flights in 1D (left panel) and 2D (right panel).
 For Cartesian L\'evy flights region of non-monotonous dependence on the stability index $\alpha$ is the same like for 1D spherical L\'evy flights regardless of the dimension $d$.
}
 \label{fig:region}
\end{figure}

Left panel of Fig.~\ref{fig:hypersphere} shows dependence of the mean first passage time on the dimension $d$ of the hyper-sphere for spherically symmetric $\alpha$-stable motions.
Various curves correspond to various values of the stability index $\alpha$.
From simulations, see top left panel of Fig.~\ref{fig:hypersphere}, it looks that for the hyper-sphere
\begin{equation}
 \Tau \propto \frac{1}{d^{\alpha/2}}.
 \label{eq:scaling}
\end{equation}
This type of scaling can be deducted from the following reasoning.
First, let us recall some properties of 1D $\alpha$-stable densities.
L\'evy distribution (with $\alpha<2$) are characterized by the infinite second moment
\begin{equation}
\langle x^2 \rangle = \int_{-\infty}^{\infty} x^2 l_{\alpha}(x) dx = \infty,
\end{equation}
where $l_{\alpha}$ is the symmetric 1D $\alpha$-stable density with the characteristic function given by Eq.~(\ref{eq:characteristic1d}).
Nevertheless, in practical realizations empirical second moment grows like a power-law, due to effective cut-off of the distribution support.
A nice explanation of this fact comes from Bouchaud and Georges \cite{bouchaud1990} and is repeated in \cite{dybiec2009h}.
Among $N$ performed jumps there is the largest one, let say $l_c(N)$, whose length grows like
\begin{equation}
l_c(N)\propto N^{1/\alpha}.
\label{eq:threshold}
\end{equation}
Eq.~(\ref{eq:threshold}) gives effective threshold for the support of distribution.
Using Eq.~(\ref{eq:threshold}) and an asymptotic form of symmetric $\alpha$-stable densities
$
l_{\alpha}(x) \propto |x|^{-(1+\alpha)},
\label{eq:asymptotics}
$
it is possible to estimate $\langle x^2 \rangle$ as
\begin{equation}
\langle x^2 \rangle \approx \int^{l_c}x^2 l_{\alpha,\beta}(x) dx\approx \left( N^{1/\alpha} \right)^{2-\alpha}=N^{2/\alpha-1}.
\end{equation}
After $N$ jumps
\begin{equation}
\langle x^2 \rangle_N=N\langle x^2 \rangle \propto N^{2/\alpha} \propto t^{2/\alpha},
\label{eq:variancescaling}
\end{equation}
because jumps are performed every $\Delta t$.
Consequently, for L\'evy flights (sample based) standard deviation grows like a power law with the number of jumps $N$ (time $t$).

Escape from the interval takes place when characteristic width of distribution of position $x$ is equal to the interval half-width  $R$. More precisely when
\begin{equation}
 R^2 \approx \langle x^2 \rangle.
\end{equation}
In $d$ dimensions instead of $\langle x^2 \rangle$ one needs to calculate $\langle \boldsymbol{x}^2 \rangle$. Assuming that jumps along axes are independent
\begin{equation}
 \langle \boldsymbol{x}^2 \rangle = \langle x_1^2 \rangle+ \dots + \langle x_d^2 \rangle = d \times \langle x^2 \rangle \propto d \times t^{2/\alpha}.
\end{equation}
A random walker escapes the hyper-sphere when
$
 R^2 \approx d \times t^{2/\alpha}.
$
Therefore, the escape time scales as
\begin{equation}
 \Tau \propto \frac{R^\alpha}{d^{\alpha/2}}.
 \label{eq:scaling-reas}
\end{equation}
The above considerations assume that jumps along all axes are independent, what is not the case of general multi-variate $\alpha$-stable densities with $\alpha<2$, see \cite{samorodnitsky1994}. Nevertheless, the scaling~(\ref{eq:scaling-reas}) nicely approximates all simulations performed, see left bottom panel of Fig.~\ref{fig:hypersphere}, which presents
 the ratio of MFPTs for $d$-dimensional hyper-sphere and 1D sphere (interval), i.e. $\langle \tau \rangle / \langle \tau \rangle_{d=1} \times d^{\alpha/2}$. 
This can be further confirmed by fits and exact formula (\ref{eq:general-mfpt}) which can be approximated by $d^{-\alpha/2}$ scaling predicted by Eq.~(\ref{eq:scaling-reas}).

In the straight forward manner, the scaling given by Eq.~(\ref{eq:scaling-reas}) can be determined from the general formula for the first passage time for a symmetric $\alpha$-stable process from a $d$ dimensional hyper sphere
\cite{blumenthal1961, *getoor1961, *kac1950distribution, *widom1961stable, *kesten1961random}
\begin{equation}
\Tau \propto  \frac{\Gamma\left[ \frac{d}{2} \right]}{ 2^\alpha \sigma^\alpha \Gamma\left[ 1 + \frac{\alpha}{2} \right]  \Gamma\left[ \frac{d+\alpha}{2} \right]} R^\alpha.
\label{eq:general-mfpt}
\end{equation}
Eq.~(\ref{eq:general-mfpt}) corroborates 1D motivation presented in Sec.~\ref{sec:1d} and
for $d=1$ it reduces to Eq.~(\ref{eq:mfpt-1d}).
Finally, from Eq.~(\ref{eq:general-mfpt}) one immediately concludes that the mean first passage time scales as $\Tau \propto \sigma^{-\alpha}$.
Such a scaling holds not only for the hyper-sphere but also for all considered domains of motion (results not shown) what is the consequence of the dimensional analysis.
Moreover, using Eq.~(\ref{eq:general-mfpt}) it is possible to estimate the value of the stability index $\alpha$ resulting in the maximal value of the MFPT as a function of hyper sphere radius $R$ and scale parameter $\sigma$, see Fig.~\ref{fig:region}.

Figure~\ref{fig:region} shows the value of the stability index $\alpha$ resulting in the maximal value of the MFPT as a function of the sphere diameter $R$ and scale parameter $\sigma$ for $d=1$ (left column) and $d=2$ (right column). With the increasing hyper-sphere dimension $d$ the region of non-monotonous dependence of the MFPT on the stability index $\alpha$ decreases in comparison to less dimensional systems. This demonstrates that in higher dimensions it is less likely to observe non-monotonous dependence of the MFPT on the stability index $\alpha$ than in lower dimensional systems.

The very different situation takes place for Cartesian L\'evy flights, i.e. $\alpha$-stable motions with discrete uniform spectral measures located on intersections of axes with the unit hyper-sphere, see right column of Fig.~\ref{fig:hypersphere}.
The mean first passage time scales with the dimension $d$ as
\[
 \Tau \propto \frac{1}{d},
\]
i.e. the scaling, contrary to the spherically symmetric case, is independent of the stability index $\alpha$.
Cartesian L\'evy flights escape faster from a hyper sphere of a given radius than spherical L\'evy flights, see Fig.~\ref{fig:hypersphere}, due to their anisotropy which is manifested by the preference to move along axes.
The heuristic reasoning leading to scaling given by Eq.~(\ref{eq:scaling-reas}) cannot be extended for Cartesian L\'evy flights because of their anisotropy, i.e. probability densities are no longer spherically symmetric, see bottom panel Fig.~\ref{fig:trajectories}.

Right bottom panel of Fig.~\ref{fig:hypersphere} presents ratio of MFPTs for $d$ dimensional hyper-sphere and 1D sphere (interval), i.e. $\langle \tau \rangle / \langle \tau \rangle_{d=1} \times d$, for Cartesian L\'evy flights. The rescaled mean first passage time suggests that for Cartesian L\'evy flights the mean first passage time is
\begin{equation}
 \Tau = \langle \tau \rangle_{d=1} \times \frac{1}{d} = \frac{R^\alpha}{\Gamma(1+\alpha) \sigma^\alpha} \times \frac{1}{d}.
 \label{eq:mfptclf}
\end{equation}
Therefore, for Cartesian L\'evy flights dependence on $d$ like for $\alpha=2$ is observed, see Eq.~(\ref{eq:mfpt-d-spere}).
The hypothesis given by Eq.~(\ref{eq:mfptclf}) is consistent with the general properties of multivariate $\alpha$-stable densities. In the limit of $\alpha=2$  both (isotropic) spherical and Cartesian L\'evy flights are equivalent. Both of them represent 2D white Gaussian process with independent increments resulting in the same formula for the MFPT, i.e. $\Tau = R^2/2d\sigma^2 $. 
The difference between Cartesian and spherical L\'evy flights reveals for $\alpha<2$. 
Increments along axes of Cartesian L\'evy flights stay independent while for isotropic multivariate $\alpha$-stable motions (spherical L\'evy flights) become dependent.  
The independence of components of Cartesian L\'evy flights is responsible for the scaling of the MFPT recorded in the right panel of Fig.~\ref{fig:hypersphere}.
Contrary to the Cartesian L\'evy flights, for spherical L\'evy flights different scaling on the hyper-sphere dimension $d$ originates as a consequence of dependence among increments along axes, which can be measured  by covariation \cite{samorodnitsky1994}, codifference \cite{samorodnitsky1994} or correlation cascade \cite{eliazar2007}. 
Consequently, the decrease of the stability index $\alpha$  plays slightly different role for Cartesian and spherical L\'evy flights. In both cases it changes the probability density, but only for spherical L\'evy flights it controls dependence among components of displacements which is responsible for scaling of the MFPT on $d$, compare Eq.~(\ref{eq:general-mfpt}) and Eq.~(\ref{eq:mfptclf}).
Finally, from Eq.~(\ref{eq:mfptclf}) one can conclude that the region of non-monotonous dependence of the MFPT on the stability index $\alpha$ does not depend on the dimension $d$ and is always as the one presented in the left panel of Fig.~\ref{fig:region}.

\section{Summary and Conclusions\label{sec:summary}}

White $\alpha$-stable noise provides a natural generalization of white Gaussian noise including the latter as a special limiting case of  $\alpha=2$.
Symmetric one dimensional $\alpha$-stable noises are characterized by two parameters.
Consequently, systems driven by $\alpha$-stable noise can display richer behavior than their Gaussian white noise driven counterparts.
One of such examples is an escape problem from the finite interval restricted by two absorbing boundaries.
If the escape is driven by Gaussian white noise, the mean first passage time depends on the interval width, initial position and the noise intensity. In general the MFPT decreases with the increase of the noise intensity because large noise intensity enhances probability of longer jumps.

The very different situation takes place when the escape process is driven by $\alpha$-stable noise. 
If the noise is symmetric, the mean first passage time is not only characterized by the scale parameter $\sigma$ but also by the stability index $\alpha$, which controls the tails' asymptotics.
As in the white Gaussian noise driven escape,  the MFPT decreases with the increase of scale parameter.
However, for appropriate choice of parameters, the MFPT can be non-monotonous function of the stability index $\alpha$ due to interplay between noise parameters.

Within the current manuscript noise driven escape from bounded domains has been studied in order to verify if the non-monotonous dependence of the mean first passage time on the stability index is observed in higher-dimensions for various types of multivariate $\alpha$-stable motions. As a basic setup the escape from the disk has been considered. Extensive numerical simulations have corroborated the possibility of non-monotonous dependence of the MFPT on the stability index $\alpha$ when the whole disk edge is absorbing both for spherical and Cartesian L\'evy flights.
Spherical and Cartesian L\'evy flights result in very different scaling of the MFPT with the increasing hyper-sphere dimension $d$.
For spherical L\'evy flights, with the increase of the hyper-sphere dimension $d$ the region with non-monotonous dependence of the mean first passage time on the stability index $\alpha$ decreases.
At the same time, for Cartesian L\'evy flights, this region does not depend on the dimension $d$.


In general the problem of multi-variate noise induced escape provides a challenging task due to properties of multi-variate $\alpha$-stable noises. In 1D parameters characterizing $\alpha$-stable noises have clear and intuitive interpretation. In the multi-variate case main characteristics of $\alpha$-stable noises are determined by properties of the spectral measure. Various spectral measures result in various escape scenarios. Here, we have focused on uniform continuous spectral measures leading to spherically symmetric jumps' length distributions and discrete uniform spectral measures resulting in the so called Cartesian L\'evy flights in order to elaborate the difference between various escape scenarios.



\begin{acknowledgments}
This project has been supported in part by the grant from National Science Center (2014/13/B/ST2/020140).
Computer simulations have been performed at the Academic
Computer Center Cyfronet, Akademia G\'orniczo-Hutnicza (Krak\'ow, Poland) under CPU grant
MNiSW/Zeus\_lokalnie/UJ/052/2012.

\end{acknowledgments}


\begin{thebibliography}{83}%
\makeatletter
\providecommand \@ifxundefined [1]{%
 \@ifx{#1\undefined}
}%
\providecommand \@ifnum [1]{%
 \ifnum #1\expandafter \@firstoftwo
 \else \expandafter \@secondoftwo
 \fi
}%
\providecommand \@ifx [1]{%
 \ifx #1\expandafter \@firstoftwo
 \else \expandafter \@secondoftwo
 \fi
}%
\providecommand \natexlab [1]{#1}%
\providecommand \enquote  [1]{``#1''}%
\providecommand \bibnamefont  [1]{#1}%
\providecommand \bibfnamefont [1]{#1}%
\providecommand \citenamefont [1]{#1}%
\providecommand \href@noop [0]{\@secondoftwo}%
\providecommand \href [0]{\begingroup \@sanitize@url \@href}%
\providecommand \@href[1]{\@@startlink{#1}\@@href}%
\providecommand \@@href[1]{\endgroup#1\@@endlink}%
\providecommand \@sanitize@url [0]{\catcode `\\12\catcode `\$12\catcode
  `\&12\catcode `\#12\catcode `\^12\catcode `\_12\catcode `\%12\relax}%
\providecommand \@@startlink[1]{}%
\providecommand \@@endlink[0]{}%
\providecommand \url  [0]{\begingroup\@sanitize@url \@url }%
\providecommand \@url [1]{\endgroup\@href {#1}{\urlprefix }}%
\providecommand \urlprefix  [0]{URL }%
\providecommand \Eprint [0]{\href }%
\providecommand \doibase [0]{http://dx.doi.org/}%
\providecommand \selectlanguage [0]{\@gobble}%
\providecommand \bibinfo  [0]{\@secondoftwo}%
\providecommand \bibfield  [0]{\@secondoftwo}%
\providecommand \translation [1]{[#1]}%
\providecommand \BibitemOpen [0]{}%
\providecommand \bibitemStop [0]{}%
\providecommand \bibitemNoStop [0]{.\EOS\space}%
\providecommand \EOS [0]{\spacefactor3000\relax}%
\providecommand \BibitemShut  [1]{\csname bibitem#1\endcsname}%
\let\auto@bib@innerbib\@empty
\bibitem [{\citenamefont {Montroll}\ and\ \citenamefont
  {Weiss}(1965)}]{montroll1965}%
  \BibitemOpen
  \bibfield  {author} {\bibinfo {author} {\bibfnamefont {E.~W.}\ \bibnamefont
  {Montroll}}\ and\ \bibinfo {author} {\bibfnamefont {G.~H.}\ \bibnamefont
  {Weiss}},\ }\href@noop {} {\bibfield  {journal} {\bibinfo  {journal} {J.
  Math. Phys.}\ }\textbf {\bibinfo {volume} {6}},\ \bibinfo {pages} {167}
  (\bibinfo {year} {1965})}\BibitemShut {NoStop}%
\bibitem [{\citenamefont {Montroll}\ and\ \citenamefont
  {Shlesinger}(1984)}]{montroll1984}%
  \BibitemOpen
  \bibfield  {author} {\bibinfo {author} {\bibfnamefont {E.~W.}\ \bibnamefont
  {Montroll}}\ and\ \bibinfo {author} {\bibfnamefont {M.~F.}\ \bibnamefont
  {Shlesinger}},\ }in\ \href@noop {} {\emph {\bibinfo {booktitle} {{L\'evy}
  processes: Theory and applications}}},\ \bibinfo {editor} {edited by\
  \bibinfo {editor} {\bibfnamefont {J.~L.}\ \bibnamefont {Lebowitz}}\ and\
  \bibinfo {editor} {\bibfnamefont {E.~W.}\ \bibnamefont {Montroll}}}\
  (\bibinfo  {publisher} {North Holland},\ \bibinfo {address} {Amsterdam},\
  \bibinfo {year} {1984})\ pp.\ \bibinfo {pages} {1--121}\BibitemShut {NoStop}%
\bibitem [{\citenamefont {Metzler}\ and\ \citenamefont
  {Klafter}(2000)}]{metzler2000}%
  \BibitemOpen
  \bibfield  {author} {\bibinfo {author} {\bibfnamefont {R.}~\bibnamefont
  {Metzler}}\ and\ \bibinfo {author} {\bibfnamefont {J.}~\bibnamefont
  {Klafter}},\ }\href@noop {} {\bibfield  {journal} {\bibinfo  {journal} {Phys.
  Rep.}\ }\textbf {\bibinfo {volume} {339}},\ \bibinfo {pages} {1} (\bibinfo
  {year} {2000})}\BibitemShut {NoStop}%
\bibitem [{\citenamefont {Shlesinger}(1983)}]{shlesinger1983}%
  \BibitemOpen
  \bibfield  {author} {\bibinfo {author} {\bibfnamefont {M.~F.}\ \bibnamefont
  {Shlesinger}},\ }\href@noop {} {\bibfield  {journal} {\bibinfo  {journal} {J.
  Chem. Phys.}\ }\textbf {\bibinfo {volume} {78}},\ \bibinfo {pages} {416}
  (\bibinfo {year} {1983})}\BibitemShut {NoStop}%
\bibitem [{\citenamefont {Metzler}\ and\ \citenamefont
  {Klafter}(2004)}]{metzler2004}%
  \BibitemOpen
  \bibfield  {author} {\bibinfo {author} {\bibfnamefont {R.}~\bibnamefont
  {Metzler}}\ and\ \bibinfo {author} {\bibfnamefont {J.}~\bibnamefont
  {Klafter}},\ }\href@noop {} {\bibfield  {journal} {\bibinfo  {journal} {J.
  Phys. A: Math. Gen.}\ }\textbf {\bibinfo {volume} {37}},\ \bibinfo {pages}
  {R161} (\bibinfo {year} {2004})}\BibitemShut {NoStop}%
\bibitem [{\citenamefont {Bartumeus}\ \emph {et~al.}(2005)\citenamefont
  {Bartumeus}, \citenamefont {da~Luz}, \citenamefont {Viswanathan},\ and\
  \citenamefont {Catalan}}]{Bartumeus2005}%
  \BibitemOpen
  \bibfield  {author} {\bibinfo {author} {\bibfnamefont {F.}~\bibnamefont
  {Bartumeus}}, \bibinfo {author} {\bibfnamefont {M.~G.~E.}\ \bibnamefont
  {da~Luz}}, \bibinfo {author} {\bibfnamefont {G.~M.}\ \bibnamefont
  {Viswanathan}}, \ and\ \bibinfo {author} {\bibfnamefont {J.}~\bibnamefont
  {Catalan}},\ }\href@noop {} {\bibfield  {journal} {\bibinfo  {journal}
  {Ecology}\ }\textbf {\bibinfo {volume} {86}},\ \bibinfo {pages} {3078}
  (\bibinfo {year} {2005})}\BibitemShut {NoStop}%
\bibitem [{\citenamefont {Esposito}\ and\ \citenamefont
  {Lindenberg}(2008)}]{esposito2008}%
  \BibitemOpen
  \bibfield  {author} {\bibinfo {author} {\bibfnamefont {M.}~\bibnamefont
  {Esposito}}\ and\ \bibinfo {author} {\bibfnamefont {K.}~\bibnamefont
  {Lindenberg}},\ }\href {\doibase 10.1103/PhysRevE.77.051119} {\bibfield
  {journal} {\bibinfo  {journal} {Phys. Rev. E}\ }\textbf {\bibinfo {volume}
  {77}},\ \bibinfo {pages} {051119} (\bibinfo {year} {2008})}\BibitemShut
  {NoStop}%
\bibitem [{\citenamefont {Scalas}(2006)}]{scalas2006}%
  \BibitemOpen
  \bibfield  {author} {\bibinfo {author} {\bibfnamefont {E.}~\bibnamefont
  {Scalas}},\ }\href@noop {} {\bibfield  {journal} {\bibinfo  {journal}
  {Physica A}\ }\textbf {\bibinfo {volume} {362}},\ \bibinfo {pages} {225}
  (\bibinfo {year} {2006})}\BibitemShut {NoStop}%
\bibitem [{\citenamefont {Abad}\ \emph {et~al.}(2010)\citenamefont {Abad},
  \citenamefont {Yuste},\ and\ \citenamefont {Lindenberg}}]{abad2010}%
  \BibitemOpen
  \bibfield  {author} {\bibinfo {author} {\bibfnamefont {E.}~\bibnamefont
  {Abad}}, \bibinfo {author} {\bibfnamefont {S.~B.}\ \bibnamefont {Yuste}}, \
  and\ \bibinfo {author} {\bibfnamefont {K.}~\bibnamefont {Lindenberg}},\
  }\href {\doibase 10.1103/PhysRevE.81.031115} {\bibfield  {journal} {\bibinfo
  {journal} {Phys. Rev. E}\ }\textbf {\bibinfo {volume} {81}},\ \bibinfo
  {pages} {031115} (\bibinfo {year} {2010})}\BibitemShut {NoStop}%
\bibitem [{\citenamefont {Bar-Haim}\ and\ \citenamefont
  {Klafter}(1998)}]{barhaim1998}%
  \BibitemOpen
  \bibfield  {author} {\bibinfo {author} {\bibfnamefont {A.}~\bibnamefont
  {Bar-Haim}}\ and\ \bibinfo {author} {\bibfnamefont {J.}~\bibnamefont
  {Klafter}},\ }\href@noop {} {\bibfield  {journal} {\bibinfo  {journal} {J.
  Chem. Phys.}\ }\textbf {\bibinfo {volume} {109}},\ \bibinfo {pages} {5187}
  (\bibinfo {year} {1998})}\BibitemShut {NoStop}%
\bibitem [{\citenamefont {Solomon}\ \emph {et~al.}(1993)\citenamefont
  {Solomon}, \citenamefont {Weeks},\ and\ \citenamefont
  {Swinney}}]{solomon1993}%
  \BibitemOpen
  \bibfield  {author} {\bibinfo {author} {\bibfnamefont {T.~H.}\ \bibnamefont
  {Solomon}}, \bibinfo {author} {\bibfnamefont {E.~R.}\ \bibnamefont {Weeks}},
  \ and\ \bibinfo {author} {\bibfnamefont {H.~L.}\ \bibnamefont {Swinney}},\
  }\href {\doibase 10.1103/PhysRevLett.71.3975} {\bibfield  {journal} {\bibinfo
   {journal} {Phys. Rev. Lett.}\ }\textbf {\bibinfo {volume} {71}},\ \bibinfo
  {pages} {3975} (\bibinfo {year} {1993})}\BibitemShut {NoStop}%
\bibitem [{\citenamefont {Solomon}\ \emph {et~al.}(1994)\citenamefont
  {Solomon}, \citenamefont {Weeks},\ and\ \citenamefont
  {Swinney}}]{solomon1994}%
  \BibitemOpen
  \bibfield  {author} {\bibinfo {author} {\bibfnamefont {T.~H.}\ \bibnamefont
  {Solomon}}, \bibinfo {author} {\bibfnamefont {E.~R.}\ \bibnamefont {Weeks}},
  \ and\ \bibinfo {author} {\bibfnamefont {H.~L.}\ \bibnamefont {Swinney}},\
  }\href@noop {} {\bibfield  {journal} {\bibinfo  {journal} {Physica D}\
  }\textbf {\bibinfo {volume} {76}},\ \bibinfo {pages} {70} (\bibinfo {year}
  {1994})}\BibitemShut {NoStop}%
\bibitem [{\citenamefont {Chechkin}\ \emph
  {et~al.}(2002{\natexlab{a}})\citenamefont {Chechkin}, \citenamefont
  {Gonchar},\ and\ \citenamefont {Szyd{\l}owski}}]{chechkin2002b}%
  \BibitemOpen
  \bibfield  {author} {\bibinfo {author} {\bibfnamefont {A.~V.}\ \bibnamefont
  {Chechkin}}, \bibinfo {author} {\bibfnamefont {V.~Y.}\ \bibnamefont
  {Gonchar}}, \ and\ \bibinfo {author} {\bibfnamefont {M.}~\bibnamefont
  {Szyd{\l}owski}},\ }\href@noop {} {\bibfield  {journal} {\bibinfo  {journal}
  {Phys. Plasmas}\ }\textbf {\bibinfo {volume} {9}},\ \bibinfo {pages} {78}
  (\bibinfo {year} {2002}{\natexlab{a}})}\BibitemShut {NoStop}%
\bibitem [{\citenamefont {Boldyrev}\ and\ \citenamefont
  {Gwinn}(2003)}]{boldyrev2003}%
  \BibitemOpen
  \bibfield  {author} {\bibinfo {author} {\bibfnamefont {S.}~\bibnamefont
  {Boldyrev}}\ and\ \bibinfo {author} {\bibfnamefont {C.~R.}\ \bibnamefont
  {Gwinn}},\ }\href@noop {} {\bibfield  {journal} {\bibinfo  {journal} {Phys.
  Rev. Lett.}\ }\textbf {\bibinfo {volume} {91}},\ \bibinfo {pages} {131101}
  (\bibinfo {year} {2003})}\BibitemShut {NoStop}%
\bibitem [{\citenamefont {Shlesinger}\ \emph {et~al.}(1995)\citenamefont
  {Shlesinger}, \citenamefont {Zaslavsky},\ and\ \citenamefont
  {Frisch}}]{shlesinger1995}%
  \BibitemOpen
  \bibinfo {editor} {\bibfnamefont {M.~F.}\ \bibnamefont {Shlesinger}},
  \bibinfo {editor} {\bibfnamefont {G.~M.}\ \bibnamefont {Zaslavsky}}, \ and\
  \bibinfo {editor} {\bibfnamefont {J.}~\bibnamefont {Frisch}},\ eds.,\
  \href@noop {} {\emph {\bibinfo {title} {{L\'evy} flights and related topics
  in physics}}}\ (\bibinfo  {publisher} {Springer Verlag},\ \bibinfo {address}
  {Berlin},\ \bibinfo {year} {1995})\BibitemShut {NoStop}%
\bibitem [{\citenamefont {Barndorff-Nielsen}\ \emph {et~al.}(2001)\citenamefont
  {Barndorff-Nielsen}, \citenamefont {Mikosch},\ and\ \citenamefont
  {Resnick}}]{nielsen2001}%
  \BibitemOpen
  \bibinfo {editor} {\bibfnamefont {O.~E.}\ \bibnamefont {Barndorff-Nielsen}},
  \bibinfo {editor} {\bibfnamefont {T.}~\bibnamefont {Mikosch}}, \ and\
  \bibinfo {editor} {\bibfnamefont {S.~I.}\ \bibnamefont {Resnick}},\ eds.,\
  \href@noop {} {\emph {\bibinfo {title} {{L\'evy} processes: Theory and
  applications}}}\ (\bibinfo  {publisher} {Birkh\"auser},\ \bibinfo {address}
  {Boston},\ \bibinfo {year} {2001})\BibitemShut {NoStop}%
\bibitem [{\citenamefont {Ditlevsen}(1999)}]{ditlevsen1999b}%
  \BibitemOpen
  \bibfield  {author} {\bibinfo {author} {\bibfnamefont {P.~D.}\ \bibnamefont
  {Ditlevsen}},\ }\href@noop {} {\bibfield  {journal} {\bibinfo  {journal}
  {Geophys. Res. Lett.}\ }\textbf {\bibinfo {volume} {26}},\ \bibinfo {pages}
  {1441} (\bibinfo {year} {1999})}\BibitemShut {NoStop}%
\bibitem [{\citenamefont {Mantegna}\ and\ \citenamefont
  {Stanley}(2000)}]{mantegna2000}%
  \BibitemOpen
  \bibfield  {author} {\bibinfo {author} {\bibfnamefont {R.~N.}\ \bibnamefont
  {Mantegna}}\ and\ \bibinfo {author} {\bibfnamefont {H.~E.}\ \bibnamefont
  {Stanley}},\ }\href@noop {} {\emph {\bibinfo {title} {An introduction to
  econophysics. Correlations and complexity in finance}}}\ (\bibinfo
  {publisher} {Cambridge University Press},\ \bibinfo {address} {Cambridge},\
  \bibinfo {year} {2000})\BibitemShut {NoStop}%
\bibitem [{\citenamefont {Brockmann}\ \emph {et~al.}(2006)\citenamefont
  {Brockmann}, \citenamefont {Hufnagel},\ and\ \citenamefont
  {Geisel}}]{brockmann2006}%
  \BibitemOpen
  \bibfield  {author} {\bibinfo {author} {\bibfnamefont {D.}~\bibnamefont
  {Brockmann}}, \bibinfo {author} {\bibfnamefont {L.}~\bibnamefont {Hufnagel}},
  \ and\ \bibinfo {author} {\bibfnamefont {T.}~\bibnamefont {Geisel}},\
  }\href@noop {} {\bibfield  {journal} {\bibinfo  {journal} {Nature (London)}\
  }\textbf {\bibinfo {volume} {439}},\ \bibinfo {pages} {462} (\bibinfo {year}
  {2006})}\BibitemShut {NoStop}%
\bibitem [{\citenamefont {Dybiec}\ \emph {et~al.}(2009)\citenamefont {Dybiec},
  \citenamefont {Kleczkowski},\ and\ \citenamefont {Gilligan}}]{dybiec2009c}%
  \BibitemOpen
  \bibfield  {author} {\bibinfo {author} {\bibfnamefont {B.}~\bibnamefont
  {Dybiec}}, \bibinfo {author} {\bibfnamefont {A.}~\bibnamefont {Kleczkowski}},
  \ and\ \bibinfo {author} {\bibfnamefont {C.~A.}\ \bibnamefont {Gilligan}},\
  }\href {\doibase 10.1098/rsif.2008.0468} {\bibfield  {journal} {\bibinfo
  {journal} {J. R. Soc. Interface}\ }\textbf {\bibinfo {volume} {6}},\ \bibinfo
  {pages} {941} (\bibinfo {year} {2009})}\BibitemShut {NoStop}%
\bibitem [{\citenamefont {Chechkin}\ \emph {et~al.}(2004)\citenamefont
  {Chechkin}, \citenamefont {Gonchar}, \citenamefont {Klafter}, \citenamefont
  {Metzler},\ and\ \citenamefont {Tanatarov}}]{chechkin2004}%
  \BibitemOpen
  \bibfield  {author} {\bibinfo {author} {\bibfnamefont {A.~V.}\ \bibnamefont
  {Chechkin}}, \bibinfo {author} {\bibfnamefont {V.~Y.}\ \bibnamefont
  {Gonchar}}, \bibinfo {author} {\bibfnamefont {J.}~\bibnamefont {Klafter}},
  \bibinfo {author} {\bibfnamefont {R.}~\bibnamefont {Metzler}}, \ and\
  \bibinfo {author} {\bibfnamefont {L.~V.}\ \bibnamefont {Tanatarov}},\
  }\href@noop {} {\bibfield  {journal} {\bibinfo  {journal} {J. Stat. Phys.}\
  }\textbf {\bibinfo {volume} {115}},\ \bibinfo {pages} {1505} (\bibinfo {year}
  {2004})}\BibitemShut {NoStop}%
\bibitem [{\citenamefont {Sokolov}\ and\ \citenamefont
  {Belik}(2003)}]{sokolov2003}%
  \BibitemOpen
  \bibfield  {author} {\bibinfo {author} {\bibfnamefont {I.~M.}\ \bibnamefont
  {Sokolov}}\ and\ \bibinfo {author} {\bibfnamefont {V.~V.}\ \bibnamefont
  {Belik}},\ }\href@noop {} {\bibfield  {journal} {\bibinfo  {journal} {Physica
  A}\ }\textbf {\bibinfo {volume} {330}},\ \bibinfo {pages} {46} (\bibinfo
  {year} {2003})}\BibitemShut {NoStop}%
\bibitem [{\citenamefont {Dubkov}\ \emph {et~al.}(2008)\citenamefont {Dubkov},
  \citenamefont {Spagnolo},\ and\ \citenamefont {Uchaikin}}]{dubkov2008}%
  \BibitemOpen
  \bibfield  {author} {\bibinfo {author} {\bibfnamefont {A.~A.}\ \bibnamefont
  {Dubkov}}, \bibinfo {author} {\bibfnamefont {B.}~\bibnamefont {Spagnolo}}, \
  and\ \bibinfo {author} {\bibfnamefont {V.~V.}\ \bibnamefont {Uchaikin}},\
  }\href@noop {} {\bibfield  {journal} {\bibinfo  {journal} {Int. J.
  Bifurcation Chaos. Appl. Sci. Eng.}\ }\textbf {\bibinfo {volume} {18}},\
  \bibinfo {pages} {2649} (\bibinfo {year} {2008})}\BibitemShut {NoStop}%
\bibitem [{\citenamefont {Rypdal}\ and\ \citenamefont
  {Rypdal}(2010)}]{rypdal2010}%
  \BibitemOpen
  \bibfield  {author} {\bibinfo {author} {\bibfnamefont {M.}~\bibnamefont
  {Rypdal}}\ and\ \bibinfo {author} {\bibfnamefont {K.}~\bibnamefont
  {Rypdal}},\ }\href {\doibase 10.1103/PhysRevLett.104.128501} {\bibfield
  {journal} {\bibinfo  {journal} {Phys. Rev. Lett.}\ }\textbf {\bibinfo
  {volume} {104}},\ \bibinfo {pages} {128501} (\bibinfo {year}
  {2010})}\BibitemShut {NoStop}%
\bibitem [{\citenamefont {Barthelemy}\ \emph {et~al.}(2008)\citenamefont
  {Barthelemy}, \citenamefont {Bertolotti},\ and\ \citenamefont
  {Wiersma}}]{barthelemy2008}%
  \BibitemOpen
  \bibfield  {author} {\bibinfo {author} {\bibfnamefont {P.}~\bibnamefont
  {Barthelemy}}, \bibinfo {author} {\bibfnamefont {J.}~\bibnamefont
  {Bertolotti}}, \ and\ \bibinfo {author} {\bibfnamefont {D.}~\bibnamefont
  {Wiersma}},\ }\href@noop {} {\bibfield  {journal} {\bibinfo  {journal}
  {Nature (London)}\ }\textbf {\bibinfo {volume} {453}},\ \bibinfo {pages}
  {495} (\bibinfo {year} {2008})}\BibitemShut {NoStop}%
\bibitem [{\citenamefont {Pasternak}\ \emph {et~al.}(2009)\citenamefont
  {Pasternak}, \citenamefont {Bartumeus},\ and\ \citenamefont
  {Grasso}}]{pasternak2009}%
  \BibitemOpen
  \bibfield  {author} {\bibinfo {author} {\bibfnamefont {Z.}~\bibnamefont
  {Pasternak}}, \bibinfo {author} {\bibfnamefont {F.}~\bibnamefont
  {Bartumeus}}, \ and\ \bibinfo {author} {\bibfnamefont {F.~W.}\ \bibnamefont
  {Grasso}},\ }\href@noop {} {\bibfield  {journal} {\bibinfo  {journal} {J.
  Phys. A: Math. Gen.}\ }\textbf {\bibinfo {volume} {42}},\ \bibinfo {pages}
  {434010} (\bibinfo {year} {2009})}\BibitemShut {NoStop}%
\bibitem [{\citenamefont {Lomholt}\ \emph {et~al.}(2005)\citenamefont
  {Lomholt}, \citenamefont {Ambj{\"o}rnsson},\ and\ \citenamefont
  {Metzler}}]{lomholt2005}%
  \BibitemOpen
  \bibfield  {author} {\bibinfo {author} {\bibfnamefont {M.~A.}\ \bibnamefont
  {Lomholt}}, \bibinfo {author} {\bibfnamefont {T.}~\bibnamefont
  {Ambj{\"o}rnsson}}, \ and\ \bibinfo {author} {\bibfnamefont {R.}~\bibnamefont
  {Metzler}},\ }\href@noop {} {\bibfield  {journal} {\bibinfo  {journal} {Phys.
  Rev. Lett.}\ }\textbf {\bibinfo {volume} {95}},\ \bibinfo {pages} {260603}
  (\bibinfo {year} {2005})}\BibitemShut {NoStop}%
\bibitem [{\citenamefont {Klages}\ \emph {et~al.}(2008)\citenamefont {Klages},
  \citenamefont {Radons},\ and\ \citenamefont {Sokolov}}]{klages2008}%
  \BibitemOpen
  \bibfield  {author} {\bibinfo {author} {\bibfnamefont {R.}~\bibnamefont
  {Klages}}, \bibinfo {author} {\bibfnamefont {G.}~\bibnamefont {Radons}}, \
  and\ \bibinfo {author} {\bibfnamefont {I.~M.}\ \bibnamefont {Sokolov}},\
  }\href@noop {} {\emph {\bibinfo {title} {Anomalous transport: Foundations and
  applications}}}\ (\bibinfo  {publisher} {Wiley-VCH},\ \bibinfo {address}
  {Weinheim},\ \bibinfo {year} {2008})\BibitemShut {NoStop}%
\bibitem [{\citenamefont {Srokowski}(2009)}]{srokowski2009b}%
  \BibitemOpen
  \bibfield  {author} {\bibinfo {author} {\bibfnamefont {T.}~\bibnamefont
  {Srokowski}},\ }\href {\doibase 10.1103/PhysRevE.79.040104} {\bibfield
  {journal} {\bibinfo  {journal} {Phys. Rev. E}\ }\textbf {\bibinfo {volume}
  {79}},\ \bibinfo {pages} {040104} (\bibinfo {year} {2009})}\BibitemShut
  {NoStop}%
\bibitem [{\citenamefont {Dubkov}\ and\ \citenamefont
  {Spagnolo}(2013)}]{dubkov2013}%
  \BibitemOpen
  \bibfield  {author} {\bibinfo {author} {\bibfnamefont {A.~A.}\ \bibnamefont
  {Dubkov}}\ and\ \bibinfo {author} {\bibfnamefont {B.}~\bibnamefont
  {Spagnolo}},\ }\href@noop {} {\bibfield  {journal} {\bibinfo  {journal} {Eur.
  Phys. J ST}\ }\textbf {\bibinfo {volume} {216}},\ \bibinfo {pages} {31}
  (\bibinfo {year} {2013})}\BibitemShut {NoStop}%
\bibitem [{\citenamefont {Janicki}\ and\ \citenamefont
  {Weron}(1994)}]{janicki1994}%
  \BibitemOpen
  \bibfield  {author} {\bibinfo {author} {\bibfnamefont {A.}~\bibnamefont
  {Janicki}}\ and\ \bibinfo {author} {\bibfnamefont {A.}~\bibnamefont
  {Weron}},\ }\href@noop {} {\emph {\bibinfo {title} {Simulation and chaotic
  behavior of $\alpha$-stable stochastic processes}}}\ (\bibinfo  {publisher}
  {Marcel Dekker},\ \bibinfo {address} {New York},\ \bibinfo {year}
  {1994})\BibitemShut {NoStop}%
\bibitem [{\citenamefont {Samorodnitsky}\ and\ \citenamefont
  {Taqqu}(1994)}]{samorodnitsky1994}%
  \BibitemOpen
  \bibfield  {author} {\bibinfo {author} {\bibfnamefont {G.}~\bibnamefont
  {Samorodnitsky}}\ and\ \bibinfo {author} {\bibfnamefont {M.~S.}\ \bibnamefont
  {Taqqu}},\ }\href@noop {} {\emph {\bibinfo {title} {Stable non-{Gaussian}
  random processes: Stochastic models with infinite variance}}}\ (\bibinfo
  {publisher} {Chapman and Hall},\ \bibinfo {address} {New York},\ \bibinfo
  {year} {1994})\BibitemShut {NoStop}%
\bibitem [{\citenamefont {Gnedenko}\ and\ \citenamefont
  {Kolmogorov}(1968)}]{gnedenko1968}%
  \BibitemOpen
  \bibfield  {author} {\bibinfo {author} {\bibfnamefont {B.~V.}\ \bibnamefont
  {Gnedenko}}\ and\ \bibinfo {author} {\bibfnamefont {A.~N.}\ \bibnamefont
  {Kolmogorov}},\ }\href@noop {} {\emph {\bibinfo {title} {Limit distributions
  for sums of independent random variables}}}\ (\bibinfo  {publisher}
  {Addison--Wesley},\ \bibinfo {address} {Reading, MA},\ \bibinfo {year}
  {1968})\BibitemShut {NoStop}%
\bibitem [{\citenamefont {Meerschaert}\ and\ \citenamefont
  {Scheffler}(2001)}]{meerschaert2001}%
  \BibitemOpen
  \bibfield  {author} {\bibinfo {author} {\bibfnamefont {M.~M.}\ \bibnamefont
  {Meerschaert}}\ and\ \bibinfo {author} {\bibfnamefont {H.-P.}\ \bibnamefont
  {Scheffler}},\ }\href@noop {} {\emph {\bibinfo {title} {Limit distributions
  for sums of independent random vectors: Heavy tails in theory and
  practice}}}\ (\bibinfo  {publisher} {John Wiley \& Sons},\ \bibinfo {address}
  {New York},\ \bibinfo {year} {2001})\BibitemShut {NoStop}%
\bibitem [{\citenamefont {Benichou}\ \emph {et~al.}(2005)\citenamefont
  {Benichou}, \citenamefont {Coppey}, \citenamefont {Moreau}, \citenamefont
  {Suet},\ and\ \citenamefont {Voituriez}}]{Benichou2005a}%
  \BibitemOpen
  \bibfield  {author} {\bibinfo {author} {\bibfnamefont {O.}~\bibnamefont
  {Benichou}}, \bibinfo {author} {\bibfnamefont {M.}~\bibnamefont {Coppey}},
  \bibinfo {author} {\bibfnamefont {M.}~\bibnamefont {Moreau}}, \bibinfo
  {author} {\bibfnamefont {P.~H.}\ \bibnamefont {Suet}}, \ and\ \bibinfo
  {author} {\bibfnamefont {R.}~\bibnamefont {Voituriez}},\ }\href@noop {}
  {\bibfield  {journal} {\bibinfo  {journal} {Europhys. Lett.}\ }\textbf
  {\bibinfo {volume} {70}},\ \bibinfo {pages} {42} (\bibinfo {year}
  {2005})}\BibitemShut {NoStop}%
\bibitem [{\citenamefont {Zoia}\ \emph {et~al.}(2007)\citenamefont {Zoia},
  \citenamefont {Rosso},\ and\ \citenamefont {Kardar}}]{zoia2007}%
  \BibitemOpen
  \bibfield  {author} {\bibinfo {author} {\bibfnamefont {A.}~\bibnamefont
  {Zoia}}, \bibinfo {author} {\bibfnamefont {A.}~\bibnamefont {Rosso}}, \ and\
  \bibinfo {author} {\bibfnamefont {M.}~\bibnamefont {Kardar}},\ }\href@noop {}
  {\bibfield  {journal} {\bibinfo  {journal} {Phys. Rev. E}\ }\textbf {\bibinfo
  {volume} {76}},\ \bibinfo {pages} {021116} (\bibinfo {year}
  {2007})}\BibitemShut {NoStop}%
\bibitem [{\citenamefont {Dybiec}(2010{\natexlab{a}})}]{dybiec2010}%
  \BibitemOpen
  \bibfield  {author} {\bibinfo {author} {\bibfnamefont {B.}~\bibnamefont
  {Dybiec}},\ }\href@noop {} {\bibfield  {journal} {\bibinfo  {journal} {J.
  Stat. Mech.}\ ,\ \bibinfo {pages} {P01011}} (\bibinfo {year}
  {2010}{\natexlab{a}})}\BibitemShut {NoStop}%
\bibitem [{\citenamefont {Majumdar}\ \emph {et~al.}(2010)\citenamefont
  {Majumdar}, \citenamefont {Rosso},\ and\ \citenamefont
  {Zoia}}]{majumdar2010}%
  \BibitemOpen
  \bibfield  {author} {\bibinfo {author} {\bibfnamefont {S.~N.}\ \bibnamefont
  {Majumdar}}, \bibinfo {author} {\bibfnamefont {A.}~\bibnamefont {Rosso}}, \
  and\ \bibinfo {author} {\bibfnamefont {A.}~\bibnamefont {Zoia}},\ }\href@noop
  {} {\bibfield  {journal} {\bibinfo  {journal} {Phys. Rev. Lett.}\ }\textbf
  {\bibinfo {volume} {104}},\ \bibinfo {pages} {020602} (\bibinfo {year}
  {2010})}\BibitemShut {NoStop}%
\bibitem [{\citenamefont {Dybiec}(2010{\natexlab{b}})}]{dybiec2010c}%
  \BibitemOpen
  \bibfield  {author} {\bibinfo {author} {\bibfnamefont {B.}~\bibnamefont
  {Dybiec}},\ }\href@noop {} {\bibfield  {journal} {\bibinfo  {journal} {Acta.
  Phys. Pol. B}\ }\textbf {\bibinfo {volume} {41}},\ \bibinfo {pages} {1127}
  (\bibinfo {year} {2010}{\natexlab{b}})}\BibitemShut {NoStop}%
\bibitem [{\citenamefont {Bertoin}(1996)}]{bertoin1996first}%
  \BibitemOpen
  \bibfield  {author} {\bibinfo {author} {\bibfnamefont {J.}~\bibnamefont
  {Bertoin}},\ }\href@noop {} {\bibfield  {journal} {\bibinfo  {journal} {Bull.
  Lond. Math. Soc.}\ }\textbf {\bibinfo {volume} {28}},\ \bibinfo {pages} {514}
  (\bibinfo {year} {1996})}\BibitemShut {NoStop}%
\bibitem [{\citenamefont {Garc{\'\i}a-Garc{\'\i}a}\ \emph
  {et~al.}(2012)\citenamefont {Garc{\'\i}a-Garc{\'\i}a}, \citenamefont
  {Rosso},\ and\ \citenamefont {Schehr}}]{garcia2012}%
  \BibitemOpen
  \bibfield  {author} {\bibinfo {author} {\bibfnamefont {R.}~\bibnamefont
  {Garc{\'\i}a-Garc{\'\i}a}}, \bibinfo {author} {\bibfnamefont
  {A.}~\bibnamefont {Rosso}}, \ and\ \bibinfo {author} {\bibfnamefont
  {G.}~\bibnamefont {Schehr}},\ }\href@noop {} {\bibfield  {journal} {\bibinfo
  {journal} {Phys. Review E}\ }\textbf {\bibinfo {volume} {86}},\ \bibinfo
  {pages} {011101} (\bibinfo {year} {2012})}\BibitemShut {NoStop}%
\bibitem [{\citenamefont {de~Mulatier}\ \emph {et~al.}(2013)\citenamefont
  {de~Mulatier}, \citenamefont {Rosso},\ and\ \citenamefont
  {Schehr}}]{demulatier2013}%
  \BibitemOpen
  \bibfield  {author} {\bibinfo {author} {\bibfnamefont {C.}~\bibnamefont
  {de~Mulatier}}, \bibinfo {author} {\bibfnamefont {A.}~\bibnamefont {Rosso}},
  \ and\ \bibinfo {author} {\bibfnamefont {G.}~\bibnamefont {Schehr}},\
  }\href@noop {} {\bibfield  {journal} {\bibinfo  {journal} {J. Stat. Mech.}\
  }\textbf {\bibinfo {volume} {2013}},\ \bibinfo {pages} {P10006} (\bibinfo
  {year} {2013})}\BibitemShut {NoStop}%
\bibitem [{\citenamefont {Dybiec}\ \emph {et~al.}(2006)\citenamefont {Dybiec},
  \citenamefont {Gudowska-Nowak},\ and\ \citenamefont {H\"anggi}}]{dybiec2006}%
  \BibitemOpen
  \bibfield  {author} {\bibinfo {author} {\bibfnamefont {B.}~\bibnamefont
  {Dybiec}}, \bibinfo {author} {\bibfnamefont {E.}~\bibnamefont
  {Gudowska-Nowak}}, \ and\ \bibinfo {author} {\bibfnamefont {P.}~\bibnamefont
  {H\"anggi}},\ }\href@noop {} {\bibfield  {journal} {\bibinfo  {journal}
  {Phys. Rev. E}\ }\textbf {\bibinfo {volume} {73}},\ \bibinfo {pages} {046104}
  (\bibinfo {year} {2006})}\BibitemShut {NoStop}%
\bibitem [{\citenamefont {Chechkin}\ \emph
  {et~al.}(2003{\natexlab{a}})\citenamefont {Chechkin}, \citenamefont
  {Metzler}, \citenamefont {Gonchar}, \citenamefont {Klafter},\ and\
  \citenamefont {Tanatarov}}]{chechkin2003b}%
  \BibitemOpen
  \bibfield  {author} {\bibinfo {author} {\bibfnamefont {A.~V.}\ \bibnamefont
  {Chechkin}}, \bibinfo {author} {\bibfnamefont {R.}~\bibnamefont {Metzler}},
  \bibinfo {author} {\bibfnamefont {V.~Y.}\ \bibnamefont {Gonchar}}, \bibinfo
  {author} {\bibfnamefont {J.}~\bibnamefont {Klafter}}, \ and\ \bibinfo
  {author} {\bibfnamefont {L.~V.}\ \bibnamefont {Tanatarov}},\ }\href@noop {}
  {\bibfield  {journal} {\bibinfo  {journal} {J. Phys. A: Math. Gen.}\ }\textbf
  {\bibinfo {volume} {36}},\ \bibinfo {pages} {L537} (\bibinfo {year}
  {2003}{\natexlab{a}})}\BibitemShut {NoStop}%
\bibitem [{\citenamefont {Koren}\ \emph
  {et~al.}(2007{\natexlab{a}})\citenamefont {Koren}, \citenamefont {Lomholt},
  \citenamefont {Chechkin}, \citenamefont {Klafter},\ and\ \citenamefont
  {Metzler}}]{koren2007}%
  \BibitemOpen
  \bibfield  {author} {\bibinfo {author} {\bibfnamefont {T.}~\bibnamefont
  {Koren}}, \bibinfo {author} {\bibfnamefont {M.~A.}\ \bibnamefont {Lomholt}},
  \bibinfo {author} {\bibfnamefont {A.~V.}\ \bibnamefont {Chechkin}}, \bibinfo
  {author} {\bibfnamefont {J.}~\bibnamefont {Klafter}}, \ and\ \bibinfo
  {author} {\bibfnamefont {R.}~\bibnamefont {Metzler}},\ }\href@noop {}
  {\bibfield  {journal} {\bibinfo  {journal} {Phys. Rev. Lett.}\ }\textbf
  {\bibinfo {volume} {99}},\ \bibinfo {pages} {160602} (\bibinfo {year}
  {2007}{\natexlab{a}})}\BibitemShut {NoStop}%
\bibitem [{\citenamefont {Koren}\ \emph
  {et~al.}(2007{\natexlab{b}})\citenamefont {Koren}, \citenamefont {Chechkin},\
  and\ \citenamefont {Klafter}}]{koren2007b}%
  \BibitemOpen
  \bibfield  {author} {\bibinfo {author} {\bibfnamefont {T.}~\bibnamefont
  {Koren}}, \bibinfo {author} {\bibfnamefont {A.~V.}\ \bibnamefont {Chechkin}},
  \ and\ \bibinfo {author} {\bibfnamefont {J.}~\bibnamefont {Klafter}},\
  }\href@noop {} {\bibfield  {journal} {\bibinfo  {journal} {Physica A}\
  }\textbf {\bibinfo {volume} {379}},\ \bibinfo {pages} {10} (\bibinfo {year}
  {2007}{\natexlab{b}})}\BibitemShut {NoStop}%
\bibitem [{\citenamefont {Chechkin}\ \emph
  {et~al.}(2002{\natexlab{b}})\citenamefont {Chechkin}, \citenamefont
  {Klafter}, \citenamefont {Gonchar}, \citenamefont {Metzler},\ and\
  \citenamefont {Tanatarov}}]{chechkin2002}%
  \BibitemOpen
  \bibfield  {author} {\bibinfo {author} {\bibfnamefont {A.~V.}\ \bibnamefont
  {Chechkin}}, \bibinfo {author} {\bibfnamefont {J.}~\bibnamefont {Klafter}},
  \bibinfo {author} {\bibfnamefont {V.~Y.}\ \bibnamefont {Gonchar}}, \bibinfo
  {author} {\bibfnamefont {R.}~\bibnamefont {Metzler}}, \ and\ \bibinfo
  {author} {\bibfnamefont {L.~V.}\ \bibnamefont {Tanatarov}},\ }\href@noop {}
  {\bibfield  {journal} {\bibinfo  {journal} {Chem. Phys.}\ }\textbf {\bibinfo
  {volume} {284}},\ \bibinfo {pages} {233} (\bibinfo {year}
  {2002}{\natexlab{b}})}\BibitemShut {NoStop}%
\bibitem [{\citenamefont {Chechkin}\ \emph
  {et~al.}(2003{\natexlab{b}})\citenamefont {Chechkin}, \citenamefont
  {Klafter}, \citenamefont {Gonchar}, \citenamefont {Metzler},\ and\
  \citenamefont {Tanatarov}}]{chechkin2003}%
  \BibitemOpen
  \bibfield  {author} {\bibinfo {author} {\bibfnamefont {A.~V.}\ \bibnamefont
  {Chechkin}}, \bibinfo {author} {\bibfnamefont {J.}~\bibnamefont {Klafter}},
  \bibinfo {author} {\bibfnamefont {V.~Y.}\ \bibnamefont {Gonchar}}, \bibinfo
  {author} {\bibfnamefont {R.}~\bibnamefont {Metzler}}, \ and\ \bibinfo
  {author} {\bibfnamefont {L.~V.}\ \bibnamefont {Tanatarov}},\ }\href@noop {}
  {\bibfield  {journal} {\bibinfo  {journal} {Phys. Rev. E}\ }\textbf {\bibinfo
  {volume} {67}},\ \bibinfo {pages} {010102(R)} (\bibinfo {year}
  {2003}{\natexlab{b}})}\BibitemShut {NoStop}%
\bibitem [{\citenamefont {Dybiec}\ \emph {et~al.}(2007)\citenamefont {Dybiec},
  \citenamefont {Gudowska-Nowak},\ and\ \citenamefont {Sokolov}}]{dybiec2007d}%
  \BibitemOpen
  \bibfield  {author} {\bibinfo {author} {\bibfnamefont {B.}~\bibnamefont
  {Dybiec}}, \bibinfo {author} {\bibfnamefont {E.}~\bibnamefont
  {Gudowska-Nowak}}, \ and\ \bibinfo {author} {\bibfnamefont {I.~M.}\
  \bibnamefont {Sokolov}},\ }\href@noop {} {\bibfield  {journal} {\bibinfo
  {journal} {Phys. Rev. E}\ }\textbf {\bibinfo {volume} {76}},\ \bibinfo
  {pages} {041122} (\bibinfo {year} {2007})}\BibitemShut {NoStop}%
\bibitem [{\citenamefont {Srokowski}(2010)}]{srokowski2010}%
  \BibitemOpen
  \bibfield  {author} {\bibinfo {author} {\bibfnamefont {T.}~\bibnamefont
  {Srokowski}},\ }\href {\doibase 10.1103/PhysRevE.81.051110} {\bibfield
  {journal} {\bibinfo  {journal} {Phys. Rev. E}\ }\textbf {\bibinfo {volume}
  {81}},\ \bibinfo {pages} {051110} (\bibinfo {year} {2010})}\BibitemShut
  {NoStop}%
\bibitem [{\citenamefont {Dubkov}\ and\ \citenamefont
  {Spagnolo}(2007)}]{dubkov2007}%
  \BibitemOpen
  \bibfield  {author} {\bibinfo {author} {\bibfnamefont {A.~A.}\ \bibnamefont
  {Dubkov}}\ and\ \bibinfo {author} {\bibfnamefont {B.}~\bibnamefont
  {Spagnolo}},\ }\href@noop {} {\bibfield  {journal} {\bibinfo  {journal} {Acta
  Phys. Pol. B}\ }\textbf {\bibinfo {volume} {38}},\ \bibinfo {pages} {1745}
  (\bibinfo {year} {2007})}\BibitemShut {NoStop}%
\bibitem [{\citenamefont {Sliusarenko}\ \emph {et~al.}(2013)\citenamefont
  {Sliusarenko}, \citenamefont {Surkov}, \citenamefont {Gonchar},\ and\
  \citenamefont {Chechkin}}]{sliusarenko2012}%
  \BibitemOpen
  \bibfield  {author} {\bibinfo {author} {\bibfnamefont {O.~Y.}\ \bibnamefont
  {Sliusarenko}}, \bibinfo {author} {\bibfnamefont {D.~A.}\ \bibnamefont
  {Surkov}}, \bibinfo {author} {\bibfnamefont {V.~Y.}\ \bibnamefont {Gonchar}},
  \ and\ \bibinfo {author} {\bibfnamefont {A.~V.}\ \bibnamefont {Chechkin}},\
  }\href@noop {} {\bibfield  {journal} {\bibinfo  {journal} {Eur. Phys. J ST}\
  }\textbf {\bibinfo {volume} {216}},\ \bibinfo {pages} {133} (\bibinfo {year}
  {2013})}\BibitemShut {NoStop}%
\bibitem [{\citenamefont {Teuerle}\ and\ \citenamefont
  {Jurlewicz}(2009)}]{teuerle2009}%
  \BibitemOpen
  \bibfield  {author} {\bibinfo {author} {\bibfnamefont {M.}~\bibnamefont
  {Teuerle}}\ and\ \bibinfo {author} {\bibfnamefont {A.}~\bibnamefont
  {Jurlewicz}},\ }\href@noop {} {\bibfield  {journal} {\bibinfo  {journal}
  {Acta Phys. Pol. B}\ }\textbf {\bibinfo {volume} {40}},\ \bibinfo {pages}
  {1333} (\bibinfo {year} {2009})}\BibitemShut {NoStop}%
\bibitem [{\citenamefont {Edwards}\ \emph {et~al.}(2007)\citenamefont
  {Edwards}, \citenamefont {Phillips}, \citenamefont {Watkins}, \citenamefont
  {Freeman}, \citenamefont {Murphy}, \citenamefont {Afanasyev}, \citenamefont
  {Buldyrev}, \citenamefont {da~Luz}, \citenamefont {Raposo}, \citenamefont
  {Stanley},\ and\ \citenamefont {Viswanathan}}]{Edwards2007}%
  \BibitemOpen
  \bibfield  {author} {\bibinfo {author} {\bibfnamefont {A.~M.}\ \bibnamefont
  {Edwards}}, \bibinfo {author} {\bibfnamefont {R.~A.}\ \bibnamefont
  {Phillips}}, \bibinfo {author} {\bibfnamefont {N.~W.}\ \bibnamefont
  {Watkins}}, \bibinfo {author} {\bibfnamefont {M.~P.}\ \bibnamefont
  {Freeman}}, \bibinfo {author} {\bibfnamefont {E.~J.}\ \bibnamefont {Murphy}},
  \bibinfo {author} {\bibfnamefont {V.}~\bibnamefont {Afanasyev}}, \bibinfo
  {author} {\bibfnamefont {S.~V.}\ \bibnamefont {Buldyrev}}, \bibinfo {author}
  {\bibfnamefont {M.~G.~E.}\ \bibnamefont {da~Luz}}, \bibinfo {author}
  {\bibfnamefont {E.~P.}\ \bibnamefont {Raposo}}, \bibinfo {author}
  {\bibfnamefont {H.~E.}\ \bibnamefont {Stanley}}, \ and\ \bibinfo {author}
  {\bibfnamefont {G.~M.}\ \bibnamefont {Viswanathan}},\ }\href@noop {}
  {\bibfield  {journal} {\bibinfo  {journal} {Nature (London)}\ }\textbf
  {\bibinfo {volume} {449}},\ \bibinfo {pages} {1044} (\bibinfo {year}
  {2007})}\BibitemShut {NoStop}%
\bibitem [{\citenamefont {Blumenthal}\ \emph {et~al.}(1961)\citenamefont
  {Blumenthal}, \citenamefont {Getoor},\ and\ \citenamefont
  {Ray}}]{blumenthal1961}%
  \BibitemOpen
  \bibfield  {author} {\bibinfo {author} {\bibfnamefont {R.~M.}\ \bibnamefont
  {Blumenthal}}, \bibinfo {author} {\bibfnamefont {R.~K.}\ \bibnamefont
  {Getoor}}, \ and\ \bibinfo {author} {\bibfnamefont {D.~B.}\ \bibnamefont
  {Ray}},\ }\href@noop {} {\bibfield  {journal} {\bibinfo  {journal} {Trans.
  Am. Math. Soc.}\ }\textbf {\bibinfo {volume} {99}},\ \bibinfo {pages} {540}
  (\bibinfo {year} {1961})}\BibitemShut {NoStop}%
\bibitem [{\citenamefont {Getoor}(1961)}]{getoor1961}%
  \BibitemOpen
  \bibfield  {author} {\bibinfo {author} {\bibfnamefont {R.~K.}\ \bibnamefont
  {Getoor}},\ }\href {http://www.jstor.org/stable/1993412} {\bibfield
  {journal} {\bibinfo  {journal} {Trans. Am. Math. Soc.}\ }\textbf {\bibinfo
  {volume} {101}},\ \bibinfo {pages} {75} (\bibinfo {year} {1961})}\BibitemShut
  {NoStop}%
\bibitem [{\citenamefont {Kac}\ and\ \citenamefont
  {Pollard}(1950)}]{kac1950distribution}%
  \BibitemOpen
  \bibfield  {author} {\bibinfo {author} {\bibfnamefont {M.}~\bibnamefont
  {Kac}}\ and\ \bibinfo {author} {\bibfnamefont {H.}~\bibnamefont {Pollard}},\
  }\href@noop {} {\bibfield  {journal} {\bibinfo  {journal} {Canadian J.
  Math.}\ }\textbf {\bibinfo {volume} {2}},\ \bibinfo {pages} {375} (\bibinfo
  {year} {1950})}\BibitemShut {NoStop}%
\bibitem [{\citenamefont {Widom}(1961)}]{widom1961stable}%
  \BibitemOpen
  \bibfield  {author} {\bibinfo {author} {\bibfnamefont {H.}~\bibnamefont
  {Widom}},\ }\href@noop {} {\bibfield  {journal} {\bibinfo  {journal} {Trans.
  Am. Math. Soc.}\ }\textbf {\bibinfo {volume} {98}},\ \bibinfo {pages} {430}
  (\bibinfo {year} {1961})}\BibitemShut {NoStop}%
\bibitem [{\citenamefont {Kesten}(1961)}]{kesten1961random}%
  \BibitemOpen
  \bibfield  {author} {\bibinfo {author} {\bibfnamefont {H.}~\bibnamefont
  {Kesten}},\ }\href@noop {} {\bibfield  {journal} {\bibinfo  {journal}
  {Illinois J.Math.}\ }\textbf {\bibinfo {volume} {5}},\ \bibinfo {pages} {267}
  (\bibinfo {year} {1961})}\BibitemShut {NoStop}%
\bibitem [{\citenamefont {Redner}(2001)}]{redner2001}%
  \BibitemOpen
  \bibfield  {author} {\bibinfo {author} {\bibfnamefont {S.}~\bibnamefont
  {Redner}},\ }\href@noop {} {\emph {\bibinfo {title} {A guide to first passage
  time processes}}}\ (\bibinfo  {publisher} {Cambridge University Press},\
  \bibinfo {address} {Cambridge},\ \bibinfo {year} {2001})\BibitemShut
  {NoStop}%
\bibitem [{\citenamefont {Borodin}\ and\ \citenamefont
  {Salminen}(2002)}]{borodin2002}%
  \BibitemOpen
  \bibfield  {author} {\bibinfo {author} {\bibfnamefont {A.~N.}\ \bibnamefont
  {Borodin}}\ and\ \bibinfo {author} {\bibfnamefont {P.}~\bibnamefont
  {Salminen}},\ }\href@noop {} {\emph {\bibinfo {title} {Handbook of Brownian
  motion: facts and formulae}}}\ (\bibinfo  {publisher} {Birkh\"auser},\
  \bibinfo {address} {Bassel},\ \bibinfo {year} {2002})\BibitemShut {NoStop}%
\bibitem [{\citenamefont {Janicki}(1996)}]{janicki1996}%
  \BibitemOpen
  \bibfield  {author} {\bibinfo {author} {\bibfnamefont {A.}~\bibnamefont
  {Janicki}},\ }\href@noop {} {\emph {\bibinfo {title} {Numerical and
  statistical approximation of stochastic differential equations with
  {non-Gaussian} measures}}}\ (\bibinfo  {publisher} {Hugo Steinhaus Centre for
  Stochastic Methods},\ \bibinfo {address} {Wroc{\l}aw},\ \bibinfo {year}
  {1996})\BibitemShut {NoStop}%
\bibitem [{\citenamefont {Fogedby}(1994)}]{fogedby1994}%
  \BibitemOpen
  \bibfield  {author} {\bibinfo {author} {\bibfnamefont {H.~C.}\ \bibnamefont
  {Fogedby}},\ }\href@noop {} {\bibfield  {journal} {\bibinfo  {journal} {Phys
  Rev. E}\ }\textbf {\bibinfo {volume} {50}},\ \bibinfo {pages} {1657}
  (\bibinfo {year} {1994})}\BibitemShut {NoStop}%
\bibitem [{\citenamefont {Metzler}\ \emph {et~al.}(1999)\citenamefont
  {Metzler}, \citenamefont {Barkai},\ and\ \citenamefont
  {Klafter}}]{metzler1999}%
  \BibitemOpen
  \bibfield  {author} {\bibinfo {author} {\bibfnamefont {R.}~\bibnamefont
  {Metzler}}, \bibinfo {author} {\bibfnamefont {E.}~\bibnamefont {Barkai}}, \
  and\ \bibinfo {author} {\bibfnamefont {J.}~\bibnamefont {Klafter}},\
  }\href@noop {} {\bibfield  {journal} {\bibinfo  {journal} {Europhys. Lett.}\
  }\textbf {\bibinfo {volume} {46}},\ \bibinfo {pages} {431} (\bibinfo {year}
  {1999})}\BibitemShut {NoStop}%
\bibitem [{\citenamefont {Yanovsky}\ \emph {et~al.}(2000)\citenamefont
  {Yanovsky}, \citenamefont {Chechkin}, \citenamefont {Schertzer},\ and\
  \citenamefont {Tur}}]{yanovsky2000}%
  \BibitemOpen
  \bibfield  {author} {\bibinfo {author} {\bibfnamefont {V.~V.}\ \bibnamefont
  {Yanovsky}}, \bibinfo {author} {\bibfnamefont {A.~V.}\ \bibnamefont
  {Chechkin}}, \bibinfo {author} {\bibfnamefont {D.}~\bibnamefont {Schertzer}},
  \ and\ \bibinfo {author} {\bibfnamefont {A.~V.}\ \bibnamefont {Tur}},\
  }\href@noop {} {\bibfield  {journal} {\bibinfo  {journal} {Physica A}\
  }\textbf {\bibinfo {volume} {282}},\ \bibinfo {pages} {13} (\bibinfo {year}
  {2000})}\BibitemShut {NoStop}%
\bibitem [{\citenamefont {Schertzer}\ \emph {et~al.}(2001)\citenamefont
  {Schertzer}, \citenamefont {Larchev\^eque}, \citenamefont {Duan},
  \citenamefont {Yanowsky},\ and\ \citenamefont {Lovejoy}}]{schertzer2001}%
  \BibitemOpen
  \bibfield  {author} {\bibinfo {author} {\bibfnamefont {D.}~\bibnamefont
  {Schertzer}}, \bibinfo {author} {\bibfnamefont {M.}~\bibnamefont
  {Larchev\^eque}}, \bibinfo {author} {\bibfnamefont {J.}~\bibnamefont {Duan}},
  \bibinfo {author} {\bibfnamefont {V.~V.}\ \bibnamefont {Yanowsky}}, \ and\
  \bibinfo {author} {\bibfnamefont {S.}~\bibnamefont {Lovejoy}},\ }\href@noop
  {} {\bibfield  {journal} {\bibinfo  {journal} {J. Math. Phys.}\ }\textbf
  {\bibinfo {volume} {42}},\ \bibinfo {pages} {200} (\bibinfo {year}
  {2001})}\BibitemShut {NoStop}%
\bibitem [{\citenamefont {Cox}\ and\ \citenamefont {Miller}(1965)}]{cox1965}%
  \BibitemOpen
  \bibfield  {author} {\bibinfo {author} {\bibfnamefont {D.~R.}\ \bibnamefont
  {Cox}}\ and\ \bibinfo {author} {\bibfnamefont {H.~D.}\ \bibnamefont
  {Miller}},\ }\href@noop {} {\emph {\bibinfo {title} {The theory of stochastic
  processes}}}\ (\bibinfo  {publisher} {Chapman and Hall},\ \bibinfo {address}
  {London},\ \bibinfo {year} {1965})\BibitemShut {NoStop}%
\bibitem [{\citenamefont {Dybiec}\ and\ \citenamefont
  {Gudowska-Nowak}(2012)}]{dybiec2012fractional}%
  \BibitemOpen
  \bibfield  {author} {\bibinfo {author} {\bibfnamefont {B.}~\bibnamefont
  {Dybiec}}\ and\ \bibinfo {author} {\bibfnamefont {E.}~\bibnamefont
  {Gudowska-Nowak}},\ }in\ \href@noop {} {\emph {\bibinfo {booktitle}
  {Fractional dynamics: recent advances}}},\ \bibinfo {editor} {edited by\
  \bibinfo {editor} {\bibfnamefont {J.}~\bibnamefont {Klafter}}, \bibinfo
  {editor} {\bibfnamefont {S.~T.}\ \bibnamefont {Lim.}}, \ and\ \bibinfo
  {editor} {\bibfnamefont {R.}~\bibnamefont {Metzler}}}\ (\bibinfo  {publisher}
  {World Scientific Publishing},\ \bibinfo {address} {Singapore},\ \bibinfo
  {year} {2012})\ p.~\bibinfo {pages} {33}\BibitemShut {NoStop}%
\bibitem [{\citenamefont {Gardiner}(2009)}]{gardiner2009}%
  \BibitemOpen
  \bibfield  {author} {\bibinfo {author} {\bibfnamefont {C.~W.}\ \bibnamefont
  {Gardiner}},\ }\href@noop {} {\emph {\bibinfo {title} {Handbook of stochastic
  methods for physics, chemistry and natural sciences}}}\ (\bibinfo
  {publisher} {Springer Verlag},\ \bibinfo {address} {Berlin},\ \bibinfo {year}
  {2009})\BibitemShut {NoStop}%
\bibitem [{\citenamefont {Dybiec}(2010{\natexlab{c}})}]{dybiec2009g}%
  \BibitemOpen
  \bibfield  {author} {\bibinfo {author} {\bibfnamefont {B.}~\bibnamefont
  {Dybiec}},\ }\href@noop {} {\bibfield  {journal} {\bibinfo  {journal} {J.
  Stat. Mech.}\ ,\ \bibinfo {pages} {P01011}} (\bibinfo {year}
  {2010}{\natexlab{c}})}\BibitemShut {NoStop}%
\bibitem [{\citenamefont {Dybiec}\ and\ \citenamefont
  {Sokolov}(2014)}]{dybiec2014estimation}%
  \BibitemOpen
  \bibfield  {author} {\bibinfo {author} {\bibfnamefont {B.}~\bibnamefont
  {Dybiec}}\ and\ \bibinfo {author} {\bibfnamefont {I.~M.}\ \bibnamefont
  {Sokolov}},\ }\href {\doibase doi:10.1016/j.cpc.2014.10.007} {\bibfield
  {journal} {\bibinfo  {journal} {Comp. Phys. Comm.}\ }\textbf {\bibinfo
  {volume} {187}},\ \bibinfo {pages} {29} (\bibinfo {year} {2014})}\BibitemShut
  {NoStop}%
\bibitem [{\citenamefont {Teuerle}\ \emph {et~al.}(2012)\citenamefont
  {Teuerle}, \citenamefont {{\.Z}ebrowski},\ and\ \citenamefont
  {Magdziarz}}]{teuerle2012}%
  \BibitemOpen
  \bibfield  {author} {\bibinfo {author} {\bibfnamefont {M.}~\bibnamefont
  {Teuerle}}, \bibinfo {author} {\bibfnamefont {P.}~\bibnamefont
  {{\.Z}ebrowski}}, \ and\ \bibinfo {author} {\bibfnamefont {M.}~\bibnamefont
  {Magdziarz}},\ }\href@noop {} {\bibfield  {journal} {\bibinfo  {journal} {J.
  Phys. A: Math. Gen.}\ }\textbf {\bibinfo {volume} {45}},\ \bibinfo {pages}
  {385002} (\bibinfo {year} {2012})}\BibitemShut {NoStop}%
\bibitem [{\citenamefont {Chambers}\ \emph {et~al.}(1976)\citenamefont
  {Chambers}, \citenamefont {Mallows},\ and\ \citenamefont
  {Stuck}}]{chambers1976}%
  \BibitemOpen
  \bibfield  {author} {\bibinfo {author} {\bibfnamefont {J.~M.}\ \bibnamefont
  {Chambers}}, \bibinfo {author} {\bibfnamefont {C.~L.}\ \bibnamefont
  {Mallows}}, \ and\ \bibinfo {author} {\bibfnamefont {B.~W.}\ \bibnamefont
  {Stuck}},\ }\href@noop {} {\bibfield  {journal} {\bibinfo  {journal} {J.
  Amer. Statistical Assoc.}\ }\textbf {\bibinfo {volume} {71}},\ \bibinfo
  {pages} {340} (\bibinfo {year} {1976})}\BibitemShut {NoStop}%
\bibitem [{\citenamefont {Weron}(1996)}]{weron1996}%
  \BibitemOpen
  \bibfield  {author} {\bibinfo {author} {\bibfnamefont {R.}~\bibnamefont
  {Weron}},\ }\href@noop {} {\bibfield  {journal} {\bibinfo  {journal}
  {Statist. Probab. Lett.}\ }\textbf {\bibinfo {volume} {28}},\ \bibinfo
  {pages} {165} (\bibinfo {year} {1996})}\BibitemShut {NoStop}%
\bibitem [{\citenamefont {Modarres}\ and\ \citenamefont
  {Nolan}(1994)}]{modarres1994method}%
  \BibitemOpen
  \bibfield  {author} {\bibinfo {author} {\bibfnamefont {R.}~\bibnamefont
  {Modarres}}\ and\ \bibinfo {author} {\bibfnamefont {J.~P.}\ \bibnamefont
  {Nolan}},\ }\href@noop {} {\bibfield  {journal} {\bibinfo  {journal} {Comput.
  Stat.}\ }\textbf {\bibinfo {volume} {9}},\ \bibinfo {pages} {11} (\bibinfo
  {year} {1994})}\BibitemShut {NoStop}%
\bibitem [{\citenamefont {Nolan}(1998)}]{nolan1998b}%
  \BibitemOpen
  \bibfield  {author} {\bibinfo {author} {\bibfnamefont {J.~P.}\ \bibnamefont
  {Nolan}},\ }in\ \href@noop {} {\emph {\bibinfo {booktitle} {A practical guide
  to heavy tails: statistical techniques and applications}}},\ \bibinfo
  {editor} {edited by\ \bibinfo {editor} {\bibfnamefont {R.~J.}\ \bibnamefont
  {Feldman}}\ and\ \bibinfo {editor} {\bibfnamefont {M.~S.}\ \bibnamefont
  {Taqqu}}}\ (\bibinfo  {publisher} {Birkh\"auser},\ \bibinfo {address}
  {Boston},\ \bibinfo {year} {1998})\ p.\ \bibinfo {pages} {509}\BibitemShut
  {NoStop}%
\bibitem [{\citenamefont {Samko}\ \emph {et~al.}(1993)\citenamefont {Samko},
  \citenamefont {Kilbas},\ and\ \citenamefont {Marichev}}]{samko1993}%
  \BibitemOpen
  \bibfield  {author} {\bibinfo {author} {\bibfnamefont {S.~G.}\ \bibnamefont
  {Samko}}, \bibinfo {author} {\bibfnamefont {A.~A.}\ \bibnamefont {Kilbas}}, \
  and\ \bibinfo {author} {\bibfnamefont {O.~I.}\ \bibnamefont {Marichev}},\
  }\href@noop {} {\emph {\bibinfo {title} {Fractional integrals and
  derivatives. Theory and applications.}}}\ (\bibinfo  {publisher} {Gordon and
  Breach Science Publishers},\ \bibinfo {address} {Yverdon},\ \bibinfo {year}
  {1993})\BibitemShut {NoStop}%
\bibitem [{\citenamefont {Szczepaniec}\ and\ \citenamefont
  {Dybiec}(2014)}]{szczepaniec2014stationary}%
  \BibitemOpen
  \bibfield  {author} {\bibinfo {author} {\bibfnamefont {K.}~\bibnamefont
  {Szczepaniec}}\ and\ \bibinfo {author} {\bibfnamefont {B.}~\bibnamefont
  {Dybiec}},\ }\href@noop {} {\bibfield  {journal} {\bibinfo  {journal} {Phys.
  Rev. E}\ }\textbf {\bibinfo {volume} {90}},\ \bibinfo {pages} {032128}
  (\bibinfo {year} {2014})}\BibitemShut {NoStop}%
\bibitem [{\citenamefont {Vahabi}\ \emph {et~al.}(2013)\citenamefont {Vahabi},
  \citenamefont {Schulz}, \citenamefont {Shokri},\ and\ \citenamefont
  {Metzler}}]{vahabi2013}%
  \BibitemOpen
  \bibfield  {author} {\bibinfo {author} {\bibfnamefont {M.}~\bibnamefont
  {Vahabi}}, \bibinfo {author} {\bibfnamefont {J.~H.~P.}\ \bibnamefont
  {Schulz}}, \bibinfo {author} {\bibfnamefont {B.}~\bibnamefont {Shokri}}, \
  and\ \bibinfo {author} {\bibfnamefont {R.}~\bibnamefont {Metzler}},\
  }\href@noop {} {\bibfield  {journal} {\bibinfo  {journal} {Phys. Rev. E}\
  }\textbf {\bibinfo {volume} {87}},\ \bibinfo {pages} {042136} (\bibinfo
  {year} {2013})}\BibitemShut {NoStop}%
\bibitem [{\citenamefont {Chechkin}\ and\ \citenamefont
  {Gonchar}(2000)}]{chechkin2000b}%
  \BibitemOpen
  \bibfield  {author} {\bibinfo {author} {\bibfnamefont {A.~V.}\ \bibnamefont
  {Chechkin}}\ and\ \bibinfo {author} {\bibfnamefont {V.~Y.}\ \bibnamefont
  {Gonchar}},\ }\href@noop {} {\bibfield  {journal} {\bibinfo  {journal} {Open
  Sys. \& Information Dyn.}\ }\textbf {\bibinfo {volume} {7}},\ \bibinfo
  {pages} {375} (\bibinfo {year} {2000})}\BibitemShut {NoStop}%
\bibitem [{\citenamefont {Bouchaud}\ and\ \citenamefont
  {Georges}(1990)}]{bouchaud1990}%
  \BibitemOpen
  \bibfield  {author} {\bibinfo {author} {\bibfnamefont {J.~P.}\ \bibnamefont
  {Bouchaud}}\ and\ \bibinfo {author} {\bibfnamefont {A.}~\bibnamefont
  {Georges}},\ }\href@noop {} {\bibfield  {journal} {\bibinfo  {journal} {Phys.
  Rep.}\ }\textbf {\bibinfo {volume} {195}},\ \bibinfo {pages} {127} (\bibinfo
  {year} {1990})}\BibitemShut {NoStop}%
\bibitem [{\citenamefont {Dybiec}\ and\ \citenamefont
  {Gudowska-Nowak}(2009)}]{dybiec2009h}%
  \BibitemOpen
  \bibfield  {author} {\bibinfo {author} {\bibfnamefont {B.}~\bibnamefont
  {Dybiec}}\ and\ \bibinfo {author} {\bibfnamefont {E.}~\bibnamefont
  {Gudowska-Nowak}},\ }\href@noop {} {\bibfield  {journal} {\bibinfo  {journal}
  {Phys. Rev. E}\ }\textbf {\bibinfo {volume} {80}},\ \bibinfo {pages} {061122}
  (\bibinfo {year} {2009})}\BibitemShut {NoStop}%
\bibitem [{\citenamefont {Eliazar}\ and\ \citenamefont
  {Klafter}(2007)}]{eliazar2007}%
  \BibitemOpen
  \bibfield  {author} {\bibinfo {author} {\bibfnamefont {I.}~\bibnamefont
  {Eliazar}}\ and\ \bibinfo {author} {\bibfnamefont {J.}~\bibnamefont
  {Klafter}},\ }\href@noop {} {\bibfield  {journal} {\bibinfo  {journal}
  {Physica A}\ }\textbf {\bibinfo {volume} {376}},\ \bibinfo {pages} {1}
  (\bibinfo {year} {2007})}\BibitemShut {NoStop}%
\end{thebibliography}

\def\url#1{}

\end{document}